\documentclass[a4paper,11pt]{article}
\pdfoutput=1 

\usepackage{jcappub} 

\usepackage[T1]{fontenc} 

\usepackage{soul}
\usepackage{subcaption}
\usepackage{xcolor}

\title{Matter power spectrum induced by primordial magnetic fields: from the linear to the non-linear regime}

\author[a,b]{Pranjal Ralegankar,}
\author[c,a,d]{Enrico Garaldi}
\author[a,b,c,d,e]{and Matteo Viel}

\affiliation[a]{SISSA - International School for Advanced Studies, Via Bonomea 265, 34136 Trieste, Italy}
\affiliation[b]{INFN – National Institute for Nuclear Physics, Via Valerio 2, I-34127 Trieste, Italy}
\affiliation[c]{IFPU, Institute for Fundamental Physics of the Universe, via Beirut 2, 34151 Trieste, Italy}
\affiliation[d]{INAF, Osservatorio Astronomico di Trieste, Via G. B. Tiepolo 11, I-34131 Trieste, Italy}
\affiliation[e]{ICSC - Italian Research Center on High Performance Computing, Big Data and Quantum Computing, via Magnanelli 2, 40033,
	Casalecchio di Reno, Italy}

\emailAdd{pralegan@sissa.it}
\emailAdd{egaraldi@sissa.it}
\emailAdd{viel@sissa.it}

\abstract{Linear theory predicts that primordial magnetic fields (PMFs) enhance the matter power spectrum on small scales. However, the linear approximation breaks down on sufficiently small scales where PMF-induced baryon perturbations back-react onto the magnetic fields. Previous studies assumed that the baryon power spectrum would be sharply suppressed in this non-linear regime, based on arguments related to the magnetic Jeans scale. For the first time, we perform dedicated magnetohydrodynamic (MHD) simulations to investigate the transition from the linear to the non-linear regime. Our simulations confirm the expected linear behavior on large scales.  In the non-linear regime, however, we find that the dimensionless baryon power spectrum saturates to an $\mathcal{O}(1)$ value, which contrasts with previous analytical expectations. Additionally, our results show that several past studies overestimated the total matter power spectrum by orders of magnitude near the transition to non-linearity. Thus, the results presented in this work are useful to obtain more accurate constraints on PMFs from structure formation processes and/or different tracers of cosmic structures.}

\begin{document}
	\maketitle
	\flushbottom
	
	\section{Introduction}
	\label{sec:intro}
	
	Magnetic fields are a pervasive feature of the universe, observed not only in galaxies and clusters but also hinted at even in cosmic voids \cite{doi:10.1126/science.1184192, HESS:2014kkl, Finke:2015ona, VERITAS:2017gkr, AlvesBatista:2021sln, MAGIC:2022piy, Vovk:2023qfk, Tjemsland:2023hmj,Broderick:2011av}. While the origin of these magnetic fields in galaxies can often be traced to astrophysical processes such as the dynamo effect, the presence of magnetic fields in the voids raises intriguing questions. One compelling possibility is that these magnetic fields are primordial, dating back to the early moments of the universe. If true, primordial magnetic fields (PMFs) could provide a unique window into the physics of the early universe, potentially revealing insights into processes like inflation \cite{Ratra:1991bn, Turner:1987bw, Kobayashi:2014sga, Durrer:2010mq, Sharma:2018kgs} or phase transitions \cite{Quashnock:1988vs, Vachaspati:1991nm,Sigl:1996dm, Olea-Romacho:2023rhh, Yang:2021uid, Di:2020kbw}. Thus, it is imperative to find signals that can distinguish between the primordial and astrophysical origin of cosmic magnetic fields.
	
	PMFs can reveal their presence through various exotic effects in the early universe. Several magnetogenesis mechanisms, for instance, produce gravitational waves \cite{RoperPol:2022iel, RoperPol:2022hxn, Balaji:2024rvo}. The dissipation of PMFs in the primordial plasma can also lead to spectral distortions in the cosmic microwave background (CMB) \cite{Trivedi:2018ejz, Paoletti:2022gsn, Wagstaff:2015jaa, Jedamzik:1999bm} and heating in the intergalactic medium \cite{Sethi:2004pe,sethi:2009,Natwariya:2020ksr}. Moreover, PMFs can directly impact CMB anisotropies \cite{Planck:2015zrl, Paoletti:2019pdi,LiteBIRD:2024twk} and have recently been suggested as a possible solution to the Hubble tension \cite{Jedamzik:2020krr}. Among these effects, the most model-independent constraints arise from CMB anisotropies, which place an upper bound of around nano Gauss (nG) on PMF strengths. In contrast, to explain TeV blazar observations that suggest magnetic fields in cosmic voids, PMFs would need to be stronger than approximately $10^{-8}$  nG \cite{MAGIC:2022piy}. Meanwhile, Faraday rotation measurements, which also aim to detect large-scale magnetic fields, are not sensitive to such weak fields and instead place an upper limit of around nG \cite{Mtchedlidze:2024kvt}.
	
	One promising signal that can probe sub-nG PMFs comes from how PMFs impact the matter power spectrum. Specifically, the Lorentz force from PMFs induces large baryon density perturbations on small scales \cite{wasserman97, Kim:1994zh, Subramanian:1997gi}. Absence of such enhancement in the observed matter power spectrum has been used to place constraints on PMF strengths through various methods, including large-scale structure surveys \cite{Shaw:2010ea, Fedeli2012}, Lyman-$\alpha$ forest \cite{pandeysethi:2013,Chong13,khan13,montanino2017}, abundance of dwarf galaxies \cite{Sanati:2020,Sanati:2024ijt}, stellar mass density \cite{Zhang:2024yph}, and reionization \cite{Katz:2021, Tashiro:2006, Sethi:2004pe, pandeysethi:2015}. Additionally, future observations, such as those of the 21-cm signal \cite{Cruz:2023rmo,Bhaumik:2024efz} and line-intensity mapping \cite{Adi:2023doe, Schleicher:2008hc}, hold promise for further improving the sensitivity to PMFs.

	These constraints on PMFs have relied heavily on the linear theory. However, linear theory is applicable on large scales and for PMF spectra. On small scales, PMF-induced perturbations become large enough to back-react onto the magnetic fields. Previous studies assumed this back-reaction would lead to the suppression of the baryon power spectrum below a critical scale, often referred to as the "magnetic Jeans scale" \cite{Kim:1994zh, Subramanian:1997gi}. However, this assumption has not yet been fully tested using the complete set of non-linear equations. 
	
	In this study, we conduct the first MHD simulations to investigate how PMFs enhance density perturbations across both linear and non-linear scales. We do so by considering only perturbations induced by PMFs and neglecting the nearly scale-invariant LCDM perturbations.\footnote{For the evolution of PMFs in post-recombination universe for $\Lambda$CDM initial conditons but in the absence of PMF-induced matter perturbations, see \cite{Vazza:2020phq}.} Our simulations accurately reproduce the linear theory on large scales but reveal new behaviour in the non-linear regime. Specifically, we find that the baryon power spectrum does not follow the predicted suppression below the magnetic Jeans scale. Instead, the power spectrum nearly saturates to a constant $\mathcal{O}(1)$ value on small scales, challenging the narrative behind the magnetic Jeans scale. Moreover, previous studies that extrapolated the linear matter power spectrum down to the magnetic Jeans scale are found to overestimate the power spectrum by several orders of magnitude. These findings imply that constraints on PMFs derived from earlier works may need to be re-evaluated.
	
	Additionally, our simulations allow us to compute the matter power spectrum for PMFs with a Batchelor spectrum for the first time. This spectrum is motivated by magnetogenesis theories involving phase transitions \cite{Hosking:2022umv,Vachaspati:2020blt,Subramanian:2015lua}, but their influence on matter perturbations cannot be addressed through linear theory even on large scales. By providing the first numerical estimates of the PMF-induced power spectrum for a Batchelor spectrum, this study expands the scope of PMF research.
	
	It is important to note that our simulations are restricted to the post-recombination universe, focusing on large-scale perturbations that become non-linear after recombination. Perturbations on smaller scales, which become non-linear before recombination, could influence the recombination history \cite{Jedamzik:2018itu, Jedamzik:2023rfd} and the abundance of dark matter minihalos \cite{Ralegankar:2023pyx}. However, these small-scale perturbations fall outside the scope of this study.
	
	This paper is organized as follows. In section~\ref{sec:theory} we review the theory behind PMF's impact on cosmological density perturbations, focusing on results from linear theory. In section~\ref{sec:sim}, we discuss the results of our MHD simulations. In section~\ref{sec:fit}, we provide a semi-analytical framework to compute the matter power spectrum for a given spectrum of PMFs and compare our result with previous expectations. We conclude in section~\ref{sec:con}. The details of many of our computations are
	relegated to appendices. In appendix~\ref{sec:lambda_Di} we show that our simulations are largely unaffected by the exact location of the small-scale cut-off in the initial PMF power spectrum. In appendix~\ref{sec:scale_invt} we highlight the numerical issue in producing initial conditions for scale-invariant fields. Next, in appendix~\ref{sec:convergence} we compare simulations with different resolutions for a highly blue-tilted PMF spectrum.
	
	\section{Theoretical framework}\label{sec:theory}

    Our primary focus is on density perturbations on the largest scales where baryonic back-reaction onto primordial magnetic fields (PMFs) becomes significant. When studying these scales, we can simplify our analysis by focusing only on the post-recombination era, ignoring earlier evolution for two reasons. First, as density perturbations grow with time, the largest scale where baryons back-react on PMFs must transition from the linear to the non-linear regime after recombination. Second, in the linear limit, the governing equations have an attractor solution before recombination (see appendix~\ref{sec:pre_rec}), so we don't need precise initial conditions to solve the post-recombination evolution.
    Thus, we only need to consider evolution in the post-recombination universe for scales of our interest.
    
	After recombination, the fraction of ionised particles in the baryon plasma drops to order $\sim 10^{-4}$. Despite this small fraction, the dipole interactions between neutral and ions keep the two fluids tightly coupled \cite{Banerjee:2004df, Sethi:2004pe, Ralegankar:2024ekl}. Thus, the whole baryon fluid can be treated as a perfect conductor and magnetic fields evolve according to the ideal MHD equation,
	\begin{align}\label{eq:induction}
		\frac{\partial \vec{B}}{\partial t}&=\frac{1}{a}\nabla\times(\vec{v}_{\rm b}\times \vec{B}).
	\end{align}
	Here $\vec{B}$ is the comoving magnetic field, $\vec{B}=a^2\vec{B}_{\rm phys}$, with $\vec{B}_{\rm phys}$ being the physical magnetic field, $\vec{v}_{\rm b}$ is the baryon bulk velocity, $a$ is the scale factor, and $t$ is physical time. We set $a=1$ today.
	
	The motion of baryons is influenced by the Lorentz force from the PMFs. We are primarily interested in scales that are within the horizon by the time of recombination. On these scales, the influence of PMFs on baryons is accurately captured by \cite{Kim:1994zh}
	\begin{align}\label{eq:thetab}
		\frac{\partial \vec{v}_{\rm b}}{\partial t}+H\vec{v}_{\rm b}+\frac{(\vec{v}_{\rm b}\cdot\nabla)\vec{v}_{\rm b}}{a}+\frac{c_{\rm b}^2}{a}\nabla \delta_{\rm b}&=\frac{(\nabla\times\vec{B})\times\vec{B}}{4\pi a^5\rho_{\rm b}}-\frac{\nabla\phi}{a},
	\end{align}
	where $H=\frac{d\ln a}{dt}$ is the Hubble rate, $\rho_{\rm b}$ is the energy density of baryons, $c_{\rm b}$ is the baryon sound speed, $\delta=(\rho(x)-\bar{\rho})/\bar{\rho}$ is the fluid density perturbation and $\phi$ is the metric potential. The metric potential satisfies the Newtonian Poisson's equation,
	\begin{align}\label{eq:poisson}
		\nabla^2\phi&=\frac{1}{2M_{\rm pl}^2}a^2(\rho_{\rm b}\delta_{\rm b}+\rho_{\rm DM}\delta_{\rm DM}).
	\end{align}
	Here $M_{\rm Pl}=2.435\times 10^{18}$ GeV is the reduced Planck mass, and the subscript DM refers to dark matter.
	
	The evolution of density perturbation for both dark matter and baryon is given by the continuity equation,
	\begin{align}\label{eq:contb}
		\frac{\partial \delta}{\partial t}+\frac{\vec{\nabla}\cdot\vec{v}}{a}+\frac{\vec{\nabla}\cdot(\delta \vec{v})}{a}&=0.
	\end{align}
	The Euler equation for dark matter is the same as in eq.~\eqref{eq:thetab} but with $\vec{B}$ and $c_b$ set to zero.
	
	Because of spatial derivatives present in the Lorentz force, even scale-invariant magnetic fields induce larger perturbations on smaller scales. Thus, on small enough scales, we expect velocity perturbations to become non-linear, i.e. when $v_b/l\sim aH$. In this limit, the RHS in the induction equation (eq.~\eqref{eq:induction}) as well as the convective term ($(\vec{v}_{\rm b}\cdot\nabla)\vec{v}_{\rm b}$) in the baryon Euler equation (eq.~\eqref{eq:thetab}) cannot be neglected. On these small scales, one finds MHD-driven turbulence \cite{Trivedi:2018ejz, Banerjee:2004df, Jedamzik:2018itu}.
	
	One can analytically estimate the damping scale, $\lambda_{\rm D}$, below which the plasma becomes turbulent. Considering baryon flow is driven by the Lorentz force (see eq.~\eqref{eq:thetab}), we have 
	\begin{align}
		\sqrt{\langle v_{\rm b}^2\rangle} \sim \frac{1}{aH\lambda_{\rm D}}\frac{\langle\vec{B}^2\rangle a^{-4}}{4\pi\rho_{\rm b}}\equiv \frac{1}{aH\lambda_{\rm D}}v_A^2,
	\end{align}
	where $\langle..\rangle$ represent ensemble average and $v_A$ is the Alfven velocity of the plasma. Then by setting $v_{\rm b}/\lambda_{\rm D}\sim aH$ we obtain
	\begin{align}\label{eq:lambdaD_va}
		\lambda_{\rm D}\sim \frac{v_{\rm A}}{aH}\approx 0.1{\rm Mpc} \left(\frac{\sqrt{\langle B^2\rangle}}{\rm nG}\right).
	\end{align}
	Here we obtained the second relation by using present-day values of $\rho_{\rm m0}\approx 1.15\times 10^{-47} {\rm GeV}^4$ in the Hubble rate and $\rho_{\rm b0}\approx 1.64\times10^{-48} {\rm GeV}^4$ inside the Alfven speed.
	From the above two equations, one can see that the Alfven speed is the typical speed of baryons under Lorentz force, i.e. $\sqrt{\langle v_{\rm b}^2\rangle}\sim v_A$.  
	
	\subsection{Evolution in the linear limit}\label{sec:grow}
	To gain further insight into the evolution of MHD-driven perturbations, we take the linear limit and solve the evolution analytically. This linear limit is expected to be applicable on scales larger than $\lambda_{\rm D}$.
	
	In the linear limit, one can ignore the right-hand side of the induction equation and find 
	\begin{align}\label{eq:constB}
		\vec{B}^{\rm lin}={\rm constant\ in\ time},
	\end{align}
	where the super-script ``lin" highlights that the solution is only true in the linear limit. Similarly, one can ignore the non-linear terms in the Euler and continuity equations. Combining the Euler, continuity, and Poisson equations, and taking their Fourier transform, one obtains
	\begin{align}\label{eq:deltab_a}
		a^2\frac{\partial^2 \delta^{\rm lin}_{\rm b}}{\partial a^2}+a\frac{3}{2}\frac{\partial \delta^{\rm lin}_{\rm b}}{\partial a}+\frac{c_{\rm b}^2}{(aH)^2}k^2 \delta_{\rm b}^{\rm lin}-\frac{3}{2}\frac{\Omega_{\rm b}}{\Omega(a)}\delta^{\rm lin}_{\rm b}=-\left[\frac{3M_{\rm Pl}^2S_B}{\rho_{\rm m0}}\right]\frac{\Omega_{\rm m}}{\Omega(a)}+\frac{3}{2}\frac{\Omega_{\rm DM}}{\Omega(a)}\delta^{\rm lin}_{\rm DM}\\
		a^2\frac{\partial^2 \delta^{\rm lin}_{\rm DM}}{\partial a^2}+a\frac{3}{2}\frac{\partial \delta^{\rm lin}_{\rm DM}}{\partial a}-\frac{3}{2}\frac{\Omega_{\rm DM}}{\Omega(a)}\delta^{\rm lin}_{\rm DM}=\frac{3}{2}\frac{\Omega_{\rm b}}{\Omega(a)}\delta^{\rm lin}_{\rm b}.\label{eq:deltadm_a}
	\end{align}
	Here $k$ is the wave number of the Fourier mode,
	$\Omega(a)$ is defined as
	\begin{align}
		\Omega(a)\equiv \Omega_{\rm m}(1+a_{\rm eq}/a)+\Omega_{\Lambda}a^3,
	\end{align}
	$\Omega_{\rm b}=0.0482$ is the baryon energy density fraction today, $\Omega_{\rm m}=\Omega_{\rm b}+\Omega_{\rm DM}=0.308$ is the total matter fraction, $\Omega_{\rm DM}$ is the dark matter fraction, $\Omega_{\Lambda}=0.691$ is the dark energy fraction, $a_{\rm eq}=2.94\times 10^{-4}$ is the matter-radiation equality \cite{Planck:2018jri}, $\rho_{\rm m0}$ is the total matter energy density today, and $S_B$ is the source term from magnetic fields defined as
	\begin{align}\label{eq:S0}
		S_B=\frac{\nabla\cdot[(\nabla\times\vec{B}^{\rm lin})\times\vec{B}^{\rm lin}]}{4\pi \rho_{\rm b0}}={\rm constant\ in\ time}.
	\end{align}
	Above $\rho_{\rm b0}$ is the baryon energy density today.
	
	As the above equations are ordinary differential equations, their solution can be expressed as a linear combination of the homogeneous solution determined by the inflationary initial conditions, i.e. the standard cosmological solution $[\delta^{\rm lin}]_{\Lambda CDM}$, and the inhomogeneous solution sourced by $S_B$, $[\delta^{\rm lin}]_{PMF}$. Thus we have
\begin{align}\label{eq:lcdm_pmf}
\delta^{\rm lin} = [\delta^{\rm lin}]_{PMF} + [\delta^{\rm lin}]_{\Lambda CDM}.
\end{align}
In the linear regime, these two solutions for $\delta$ are entirely independent of one another. \textit{This study focuses on PMF-induced perturbations, $[\delta^{\rm lin}]_{PMF}$, and we therefore neglect the contributions from inflationary initial conditions.}\footnote{On scales where PMFs induce non-linear density perturbations the linear superposition of $[\delta]_{PMF}$ and $[\delta]_{\Lambda CDM}$ is not strictly valid. However, since the perturbations from inflationary initial conditions are of order $\sim 0.01$ at $z\gtrsim 100$, their impact can be safely ignored on scales where PMFs induce $\sim\mathcal{O}(1)$ perturbations.}
	
	\begin{figure}
		\begin{subfigure}{0.5\textwidth}
			\includegraphics[width=1.00\textwidth]{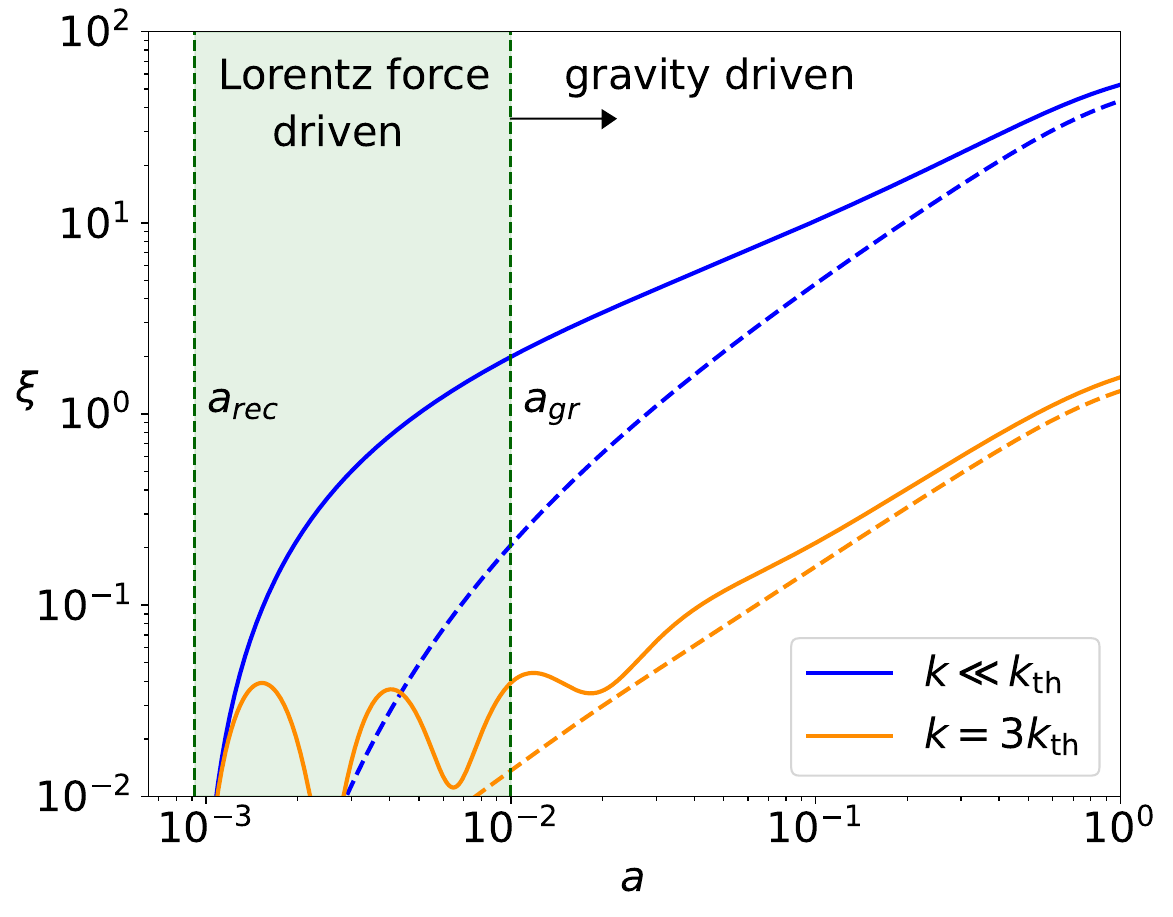}
		\end{subfigure}
		\begin{subfigure}{0.5\textwidth}
			\includegraphics[width=1.00\textwidth]{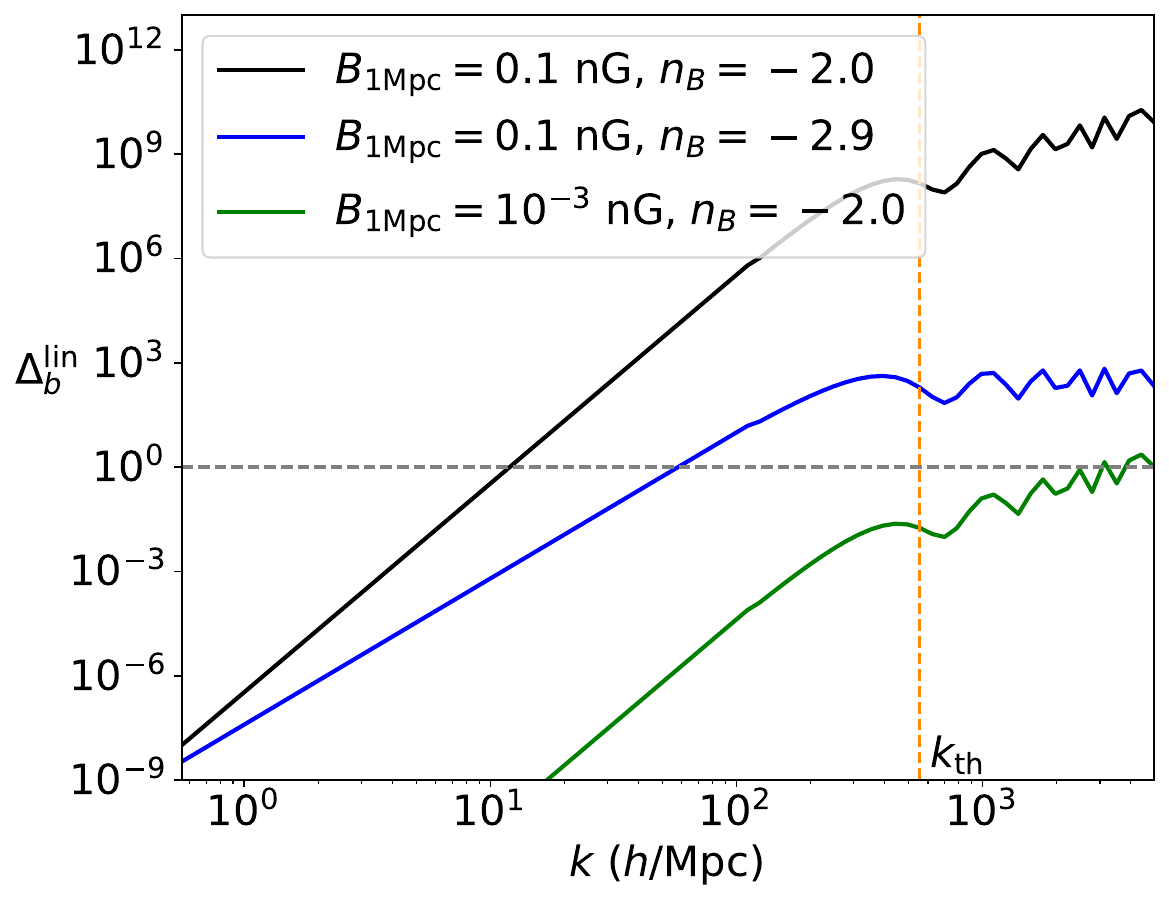}
		\end{subfigure}
		\caption{\textbf{Left}: Evolution of baryon (solid) and dark matter (dashed) density perturbations. The perturbations have been normalised with $3M_{\rm Pl}^2S_B/[a^3\rho_{\rm m}]$, which parameterises the Lorentz force. The evolution of $\xi$ is identical for all modes whose wave numbers are much smaller than the thermal Jeans scale near recombination, $k_{\rm th}$.  \textbf{Right}: Dimensionless baryon power spectrum for linearized solution at $a=0.01$ using eq.~\eqref{eq:delta_an}. The actual baryon power spectrum is expected to deviate from the shown plot of $\Delta_{\rm b}$ on scales where they are suppressed. }\label{fig:xi}
	\end{figure}
	
    The PMF-induced perturbations can be solved once the initial conditions at recombination, $a=a_{\rm rec}$, and the value of $S_B$ are specified. In Appendix~\ref{sec:pre_rec}, we demonstrate that the exact determination of the initial condition at $a_{\rm rec}$ is not critical due to the attractor nature of the solution. Therefore, for simplicity, we adopt the following trivial initial conditions:
\begin{align}\label{eq:delta_ic}
\delta^{\rm lin}(a_{\rm rec}) = 0, && \frac{\partial \delta^{\rm lin}(a_{\rm rec})}{\partial a} = 0.
\end{align}
    Note that above and in the remaining manuscript we omit the subscript ``$PMF$" from the symbol of PMF-induced density perturbations for brevity.
    
These initial conditions are strictly valid for perturbations in the linear regime. If $3M_{\rm Pl}^2 S_B / \rho_{\rm m} \gg 1$, density perturbations can become non-linear before recombination \cite{Ralegankar:2024ekl, Jedamzik:2013gua, Jedamzik:2023rfd}. However, this section focuses on idealized linear solutions and ignores the effects of non-linearities.
	
	
	With the above initial conditions, the post-recombination evolution of density perturbations can be expressed as
	\begin{align}\label{eq:xi_def}
		\delta_{\rm b}^{\rm lin}=-\xi_{\rm b}(k,a)\frac{3M_{\rm Pl}^2}{\rho_{\rm m0}}S_B(k) && \delta_{\rm DM}^{\rm lin}=-\xi_{\rm DM}(k,a)\frac{3M_{\rm Pl}^2}{\rho_{\rm m0}}S_B(k).
	\end{align}
	Here $\xi_{\rm b}$ and $\xi_{\rm DM}$ are dimensionless factors that contain the time-evolution of density perturbations and are independent of the strength of PMFs.
	
	The only scale dependence in $\xi$ comes from the thermal Jeans scale,
	\begin{align}
		\lambda_{\rm th}=k^{-1}_{\rm th}=\sqrt{\frac{2\Omega_m}{3\Omega_b}}\frac{c_b}{aH}.
	\end{align}
	The thermal Jeans scale is roughly constant post-recombination until baryons thermally decouple from photons, $z\sim 200$. After thermal decoupling the thermal Jeans scale decays as $1/\sqrt{a}$.
	
	For wave numbers $k\ll k_{\rm th}(a_{\rm rec})$, one can ignore the contribution from baryon thermal pressure and $\xi$ becomes completely scale invariant. Here, the initial behaviour of $\xi_b$ near recombination is determined by the Lorentz force. One can see this by taking the small $\delta_{\rm b}$ limit, $\delta\ll 3M_{\rm Pl}^2S_B/\rho_{\rm m0}$, where the terms with $\Omega$ in eq.~\eqref{eq:deltab_a} can be ignored. Consequently, one obtains a logarithmic growth for $\xi_{\rm b}$,
	\begin{align}\label{eq:log_xi}
		\xi_{\rm b}\approx 2\log(a/a_{\rm rec}). 
	\end{align}
	This logarithmic growth continues until self-gravity from baryons becomes important, i.e. when $\frac{3}{2}\frac{\Omega_{\rm b}}{\Omega_{\rm m}}\delta_{\rm b}\sim \frac{3M_{\rm Pl}^2S_B}{\rho_{\rm m0}}$. Substituting the above logarithmic growth of $\delta_{\rm b}$, the scale factor when gravity takes over, $a_{\rm gr}$, is then simply given by
	\begin{align}
		a_{\rm gr}\sim a_{\rm rec}\exp\left(\frac{\Omega_{\rm m}}{3\Omega_{\rm b}}\right)\approx 0.01.
	\end{align}
	After this point, gravity causes a polynomial growth in $\xi_b\propto a^m$, with the exponent $m$ eventually reaching unity as $\delta_{\rm DM}^{\rm lin}$ catches up with $\delta_{\rm b}^{\rm lin}$. The above-described behaviour can be seen in the left panel of Fig.~\ref{fig:xi}, where we show the evolution of $\xi$ after numerically solving eqs.~\eqref{eq:deltab_a} and \eqref{eq:deltadm_a} (blue lines, see also \cite{Shibusawa:2014fva, Ralegankar:2024ekl}). 
	
	Now let us consider the evolution of $\xi$ for $k\gg k_{\rm th}(a_{\rm rec})$, i.e. the orange lines in the left panel of figure~\ref{fig:xi}. Here too the very initial evolution of $\xi_b$ is given by the same logarithmic evolution as in eq.~\eqref{eq:log_xi}. However, once the baryon thermal pressure becomes of the order of the Lorentz force, $\frac{k^2c_b^2}{(aH)^2}\delta_{\rm b}\sim \frac{3M_{\rm Pl}^2S_B}{\rho_{\rm m0}}$, the baryon density perturbation starts to oscillate. In figure~\ref{fig:xi}, the transition happens when the orange line deviates from the blue line. Eventually, gravity overcomes the thermal pressure and baryon density perturbations again start to grow.
	
	\subsection{Power spectrum for the linear solution}
	To obtain the power spectrum of baryons and dark matter sourced by PMFs, one would first need to specify the spectrum for PMFs. In this study, we focus on non-helical PMFs, where
	\begin{align}\label{eq:def_M}
		\langle B_i(k) B^*_j(k')\rangle=(2\pi)^3\delta^3\!(k-k')\left(\delta_{ij}-\frac{k_ik_j}{k^2}\right)\frac{P_{\rm B}(k)}{2}.
	\end{align}
	Above $B_i(k)$ is the Fourier transform of $B_i(x)$, where we use the following convention for Fourier transforms: $A(k)=\int d^3x A(x) e^{ikx}$. With helical PMFs too we expect an enhancement of density perturbations on small scales. We leave their detailed investigation to future work.
	
	We consider the power spectrum of the magnetic fields to be of the form
	\begin{align}\label{eq:PB_powerlaw}
		P_{\rm B}(k)=Ak^{n_{\rm B}}e^{-k^2\lambda_{\rm D}^2},
	\end{align}
	where $n_{\rm B}$ is the spectral tilt set by the magnetogenesis scenario. We do not assume any specific model but impose $n_{\rm B} > -3$ to avoid generating a large homogeneous magnetic field. Next, $\lambda_{\rm D}$ is the cutoff scale set by turbulent plasma motions 
    (see eq.~\eqref{eq:lambdaD_va}).\footnote{Alternatively, one could imagine a 
    cutoff scale in PMFs arising directly from the inflationary magnetogenesis 
    scenario. If this primordial cutoff is larger than $\lambda_D$, the PMFs 
    never induce non-linear plasma motions and are not dynamically damped. 
    Conversely, if the primordial cutoff is smaller than $\lambda_D$, the PMFs 
    drive non-linear motions that eventually dissipate magnetic power on scales 
    below $\lambda_D$. In this work, for simplicity and minimality, we always 
    assume that the small-scale cutoff is set by plasma dynamics.} Strictly speaking, linear theory does not predict such a cutoff, since non-linear effects are needed to damp magnetic fields. Nevertheless, we include $\lambda_{\rm D}$ to stay consistent with previous works that use a similar approximation \cite{Kim:1994zh,Sethi:2004pe,Adi:2023doe,Finelli:2008xh,Paoletti:2008ck,Kunze:2021qxt}. Finally, the amplitude $A$ is determined by specifying the strength of PMFs averaged over $\lambda_{\rm Mpc}=1$ Mpc,
	\begin{align}\label{eq:B1mpc}
		B^2_{\rm 1 Mpc}\equiv\int \frac{d^3k}{(2\pi)^3}P^{\rm lin}_{\rm B}(k)e^{-k^2\lambda_{\rm Mpc}^2}= \frac{A\lambda_{\rm Mpc}^{-(3+n_{\rm B})}}{4\pi^2}\Gamma([n_{\rm B}+3]/2),
	\end{align}
	where $\Gamma$ is the Gamma function. In contrast to $B_{\rm 1Mpc}$, the total PMF strength is sensitive to small scales down to $\lambda_{\rm D}$ and is given by
	\begin{align}\label{eq:B_tot}
		\langle B^2\rangle=\int \frac{d^3k}{(2\pi)^3}P_{\rm B}(k)= \frac{A\lambda_{\rm D}^{-(3+n_{\rm B})}}{4\pi^2}\Gamma([n_{\rm B}+3]/2).
	\end{align}
	
	Having parameterised the spectrum of PMFs, we return to the evaluation of the power spectrum of matter fields in the linear limit, $P^{\rm lin}$. From eq.~\eqref{eq:xi_def}, one can see that  $P^{\rm lin}$ can be directly determined from the power spectrum of $S_B$. Taking the Fourier transform of $S_B$ given in eq.~\eqref{eq:S0} and neglecting non-gaussianities in PMFs while taking the ensemble average $\langle \delta(k) \delta^*(k')\rangle$, we obtain (see also \cite{Kim:1994zh, Adi:2023doe, Finelli:2008xh,Paoletti:2008ck,Kunze:2021qxt})
	\begin{multline}\label{eq:Pb_int}
		P^{\rm lin}(k) =\xi^2(k,a)\frac{9k^4M_{\rm Pl}^4}{8(4\pi \rho_{\rm b0}\rho_{\rm m0})^2}\int \frac{d^3q  }{(2\pi)^3}\frac{P_{B}(q)P_{B}(k-q)}{(k-q)^2}\bigg[k^2+2q^2+4\frac{(q\cdot k)^4}{k^4q^2}\\-4\frac{(q\cdot k)^2}{k^2}-4\frac{(q\cdot k)^3}{k^2q^2}+\frac{(q\cdot k)^2}{q^2}\bigg].
	\end{multline}
	For $-3<n_{\rm B}<-1.5$, the integral in eq.~\eqref{eq:Pb_int} is largely sensitive to $q\sim k$ and hence is decoupled from both the small and large scale cut-offs. Whereas for $n_{\rm B}>-1.5$, even the large scale spectrum is sensitive to the small scales where the evolution is non-linear. We discuss both these regimes separately below.
	
	\subsubsection{Linear theory for $n_{\rm B}<-1.5$}
	For these spectra, the large scales are decoupled from the small scale and one can integrate eq.~\eqref{eq:Pb_int} without any cutoffs. Consequently, we set $P_{\rm B}=Ak^{\rm n_{\rm B}}$ in eq.~\eqref{eq:Pb_int}, replace $A$ using eq.~\eqref{eq:B1mpc}, integrate over $q$, and use $\rho_{\rm m0}\approx 1.15\times 10^{-47} {\rm GeV}^4$ and $\rho_{\rm b0}\approx 1.64\times10^{-48} {\rm GeV}^4$ to obtain
	\begin{align}\label{eq:Delta_b}
		\Delta^{\rm lin}(k)\equiv\frac{k^3P^{\rm lin}(k)}{2\pi^2}\approx 0.918\times10^{-4}\xi^2(k,a) \left(\frac{k}{\rm Mpc^{-1}}\right)^{2n_{\rm B}+10} \left(\frac{B_{\rm 1 Mpc}}{\rm nG}\right)^4G_{\rm n_{\rm B}},
	\end{align}
	where $G_{\rm n_{\rm B}}(z)$ is a dimensionless number given by
    \begin{align}
        G_{n_{\rm B}}(z)=\int_0^{\infty} dx \int _{-1}^1\frac{dy}{2}x^{n_{\rm B}+2}(1+x^2-2xy)^{n_{\rm B}/2-1}
        \times \frac{\left[1+2x^2+4y^4x^2-4y^2x^2-4y^3x+y^2\right]}{\Gamma^2([n_{\rm B}+3]/2)},
    \end{align}
    and $\xi(k,a)$ are obtained after numerically solving eqs.~\eqref{eq:deltab_a} and \eqref{eq:deltadm_a} (see left panel of figure~\ref{fig:xi}).
	One can explain the steep $k$ dependence of $\Delta^{\rm lin}$ using the fact that $\delta\propto \nabla\cdot(\nabla\times B)\times B$. Consequently, $\Delta^{\rm lin}\propto k^4 \Delta_{\rm B}^2\propto k^{10+2n_{\rm B}}$.

    The above solution is only valid on large scales where $k\ll \lambda_D^{-1}$. On scales near and below $\lambda_D$, suppressed magnetic fields also lead to a suppressed matter power spectrum,
    \begin{align}\label{eq:delta_an}
        \Delta(k)=\Delta^{\rm lin}(k)E_{n_{\rm B}}(k\lambda_D),
    \end{align}
    where $E_{n_{\rm B}}(z)$ is a dimensionless function that sharply falls from unity to 0 as $z$ goes from 0 to values larger than 1. Its explicit form is given by,
	\begin{multline}
        E_{n_{\rm B}}(z)=\frac{1}{G_{n_{\rm B}}}\int_0^{\infty} dx \int _{-1}^1\frac{dy}{2}x^{n_{\rm B}+2}(1+x^2-2xy)^{n_{\rm B}/2-1}e^{-(1+2x^2-2xy)z^2}\\
        \times \frac{\left[1+2x^2+4y^4x^2-4y^2x^2-4y^3x+y^2\right]}{\Gamma^2([n_{\rm B}+3]/2)}.
    \end{multline}
    
	In the right panel of figure~\ref{fig:xi}, we show the baryon power spectrum 
    $\Delta_{\rm b}$ at $a=0.01$ for different magnetic field configurations. 
    We set $\lambda_D$ using equality in eq.~\eqref{eq:lambdaD_va}. 
    On large scales, $k \ll k_{\rm th}$ and $k \ll \lambda_D^{-1}$, 
    the power grows as $\Delta_{\rm b} \propto k^{2n_{\rm B}+10}$. 
    On smaller scales, $k \gtrsim \lambda_D^{-1}$, linear theory predicts 
    a Gaussian suppression of $\Delta_{\rm b}$.
    
    However, this Gaussian damping is unphysical — it arises from 
    extrapolating linear theory to scales where non-linear effects 
    cannot be ignored. Earlier works adopted a similar extrapolation 
    and concluded that the baryon power spectrum is suppressed on 
    small scales~\cite{Kim:1994zh,Sethi:2004pe,Adi:2023doe,
    Finelli:2008xh,Paoletti:2008ck,Kunze:2021qxt,Shaw:2010ea}. 
    These studies also found the peak of the baryon power spectrum 
    to depend on $n_B$, similar to the behavior seen in 
    the right panel of figure~\ref{fig:xi}.
    
    In contrast, as we show in Sec.~\ref{sec:sim}, our MHD simulations 
    find that for $k \gtrsim \lambda_D^{-1}$, $\Delta_{\rm b}$ does not 
    get suppressed but instead saturates to a value $\sim 0.5$. 
    Moreover, this saturation value is roughly independent of $n_B$ 
    (see figure~\ref{fig:Pk_final}).

	\subsubsection{Linear theory for $n_{\rm B}>-1.5$}
	For $n_{\rm B}>-1.5$, the integrand in eq.~\eqref{eq:Pb_int} diverges at large $q$. Consequently, the power spectrum is sensitive to the small-scale cut-off ($\lambda_{\rm D}$) determined by the non-linear motions. In other words, for $n_{\rm B}>-1.5$, the linearized solutions discussed in this section are not truly applicable even on large scales.
	
	While we cannot analytically compute the exact power spectrum on large scales, we can still obtain a qualitative understanding through analytical arguments. We start by noting that as long as density perturbations are small, $\delta\ll 1$, they would always follow $S_B$ such that
	\begin{align}
		\delta\propto \left[\frac{3M_{\rm Pl}^2}{\rho_{\rm m0}}S_B\right],
	\end{align}
	with the value of the proportionality constant being an unknown. Note that $S_B$ may not be constant on large scales, as opposed to what we considered in previous section (eq.~\eqref{eq:S0}).
	Consequently, the power spectrum of $\delta$ is given by the same expression as in eq.~\eqref{eq:Pb_int} but with unknown coefficients in the front. 
	
	For $k\ll \lambda_{\rm D}^{-1}$, the integral in eq.~\eqref{eq:Pb_int} is primarily determined by $q\sim\lambda_{\rm D}^{-1}\gg k$. Thus we simplify the integral by approximating $q\gg k$, yielding
	\begin{align}\label{eq:Pb_int2}
		P(k\ll \lambda_{\rm D}^{-1}) \propto \frac{9M_{\rm Pl}^2k^4}{4(4\pi \rho_{\rm b0}\rho_{\rm m0})^2}\int \frac{d^3q  }{(2\pi)^3}P_{\rm B}^2(q).
	\end{align}
	Using the fact that the above integral would be dominated by $q\sim \lambda_{\rm D}^{-1}$ scales and that the total PMF strength is given by the integral of $P_B$ (see eq.~\eqref{eq:B_tot}), we can rewrite the above expression as
	\begin{align}
		P(k\ll \lambda^{-1}_{\rm D}) \propto k^4\left(\frac{\langle B^2\rangle a^{-4}}{4\pi\rho_{\rm b}[aH]^2}\right)^2\lambda_{\rm D}^3\propto k^4\lambda_{\rm D}^7.
	\end{align}
	Above in the second step we replaced the term in the big brackets in terms of $\lambda_{\rm D}$ using eq.~\eqref{eq:lambdaD_va}. 
	
	The dimensionless power spectrum is then simply proportional to $(k\lambda_{\rm D})^7$. Thus, on large scales, the dimensionless matter power spectrum is expected to be proportional to $k^7$, independent of the value of $n_{\rm B}$ or $B_{\rm 1Mpc}$. Changing $B_{\rm 1Mpc}$ is only expected to shift the power spectrum to the left or right by shifting $\lambda_{\rm D}$. The exact value of the coefficient as well as the peak of the matter power spectrum can only be determined through MHD simulations.
	
	\section{MHD simulations}\label{sec:sim}
	In this section, we present the MHD simulations we have performed to quantify the precise behaviour of matter fields in the regime where baryons back-react on the PMFs. 
	These simulations were performed using the AREPO code \cite{arepo}, which models dark matter with particles and gas with an adaptive unstructured mesh. The mesh corresponds to the Voronoi tessellation generated by tracers that flow with the gas.
	
	Gravitational interactions are solved in AREPO using a hybrid tree-PM-direct summation algorithm. This algorithm uses a particle-mesh approach on the largest scales for speed, a hierarchical oct-tree-based solver \cite{Barnes&Hut86} on intermediate scales for accuracy, and direct summation at the shortest scales for even better precision. In the particle-mesh approach, a gridded density field is produced from the particle distribution and the gravitational potential is computed by solving the Poisson equation in Fourier space. We additionally employ a hierarchical time integration algorithm and randomisation of the box origin at each (full) force calculation to remove correlated force errors \cite{Gadget4}.
	
	AREPO follows the ideal MHD equations on the unstructured mesh by solving the Riemann problem at each cell interface. Fluxes are computed using Godunov’s approach \cite{Godunov1959}, with a slope-limited piecewise-linear spatial extrapolation and a half timestep. A first-order time extrapolation of the primitive variables is employed on both sides of the interface \cite{vanLeer1979} to achieve second-order accuracy. The spatial extrapolation employs the local gradient estimate technique presented in Ref. \cite{Pakmor2016}. Finally, gas cells can be split or merged to keep their mass within a factor of 2 of the value in the initial conditions. This ensures that the \textit{spatial} resolution adapts to the flow of gas, increasing in high-density regions and decreasing where little material is present. 
	
	Processes occurring below the scale of individual resolution elements require a so-called sub-grid model. In this work, we employ the IllustrisTNG galaxy formation model \cite{Weinberger2017,Pillepich2018b}. As this study primarily focuses on the growth of matter perturbations induced by PMFs and not on astrophysical outputs, the exact choice of sub-grid model should have a negligible impact on our final results. 
	

	\subsection{Setup of initial conditions}\label{sec:setup}
	The initial conditions (ICs) for the simulations are set to be at recombination, $z_{{\rm in}}=1090$ and are computed by modifying the NGenIC code \cite{NGENIC}. The ICs for the magnetic fields are set by the specified input magnetic field power spectrum. Specifically, our IC generation code first computes the vector potential $\vec{A}$ in Fourier space. For magnetic field power spectra with spectral slope $n_{\rm B}>-1.5$, the code then inverse Fourier transforms $\vec{A}$ and computes magnetic field as $\vec{B}=\vec{\nabla}\times \vec{A}$ (see Ref.~\cite{Ralegankar:2024ekl} for detailed algorithm). For $n_{\rm B}=2$, the above algorithm was found to generate incorrect ICs. Instead, accurate ICs are produced by first computing $B$ fields in Fourier space, $\vec{B}=-i\vec{k}\times \vec{A}$, and then inverse Fourier transforming to obtain $B$ fields in real space.

	\begin{table}
		\centering
		\caption{{\scshape Parameters of simulations}. From left to right, columns correspond to: the simulation name, primordial magnetic field (PMF) strength at 1 Mpc, spectral index of the PMF power spectrum, periodic box size, the number of baryon particles in the box and the median linear cell size at $z=4$; Below are listed values of some derived parameters: gravitational softening length, the Nyquist wave number, and particle masses of DM and gas. At the bottom, we list the values of the total matter energy density, baryon energy density and the dimensionless Hubble parameter (with $H_0 = h \times 100 \frac{\text{km}}{\text{Mpc}\,\text{s}}$). All simulations start from redshift $z_{{\rm in}}=1090$ and contain the same number of baryon and dark matter particles.}
		\begin{tabular}{c c c c c c} 
			\hline
			Simulation & $B_{\rm 1Mpc}$ & $n_{B}$ & $L_{{\rm box}}$ & $N_{\rm part}$ & $R_\mathrm{cell, med}$ ($z=4$)\\ 
			& [nG] & & [Mpc/$h$] &  & [Mpc/$h$] \\
			\hline\hline
			A & 0.2 & -2.0 & 8 & $256^3$ & 0.0202 \\
			B & 0.2 & -2.0 & 4  & $256^3$ & 0.0103\\
			C & 0.2 & -2.0 & 2  & $256^3$ & 0.005 \\
			D & 0.8 & -2.0 & 20  & $256^3$ & 0.0436\\
			E & $10^{-3}$ &  2.0 & 8 & $256^3$ & 0.0203\\ 
			F & $10^{-3}$ &  2.0 & 4 & $256^3$ & 0.0107\\
			G & 0.68 & -2.9 & 7 & $128^3$ & 0.0410\\
			\hline
		\end{tabular} \\
		\begin{tabular}{c}
			Cosmological parameters (Planck 2015 data \cite{Ade:2015xua})
		\end{tabular} \\
		\begin{tabular}{c c c c c c c c c} 
			\hline
			& & $\Omega_{\rm m} = 1-\Omega_{\Lambda}$ &  & $\Omega_{\rm b} $  &  &  $h$  &  &   \\
			&  & 0.308 &  & 4.82 $\times 10^{-2}$ &  &  0.678  &  &  \\
			\hline
		\end{tabular}
		\label{table:simparams}
	\end{table}
	
	The input magnetic field power spectrum is taken to be of the form
	\begin{align}\label{eq:PB_powerlaw2}
		P_{\rm B}(k,a=a_{rec})=Ak^{n_{\rm B}}e^{-k^2\lambda_{\rm Di}^2}.
	\end{align}
	where $A$ is determined through $B_{\rm 1Mpc}$ (see eq.~\eqref{eq:B1mpc}) and $\lambda_{\rm Di}$ is the viscous damping scale just before recombination \cite{Jedamzik:1996wp, Subramanian:1997gi} (see also appendix~\ref{sec:pre_rec}),
	\begin{align}\label{eq:lambda_Di}
		\lambda_{\rm Di}\sim 0.01 {\rm Mpc}\left(\frac{\sqrt{\langle B_i^2\rangle}}{\rm nG}\right).
	\end{align}
	Above, $B_i$ is the initial PMF strength right before recombination.
	
	The damping scale $\lambda_{\rm Di}$ is directly determined from $B_{\rm 1Mpc}$ and $n_{\rm B}$. This is because eq.~\eqref{eq:PB_powerlaw2} allows us to relate the total PMF strength to $B_{\rm 1Mpc}$, 
	\begin{align}\label{eq:B_B1mpc_i}
		\langle B_i^2\rangle=B_{\rm 1Mpc}^2\left(\frac{\rm Mpc}{\lambda_{\rm Di}}\right)^{n_{\rm B}+3}.
	\end{align}
	Using above expression of $\langle B_i^2\rangle$ in eq~.\eqref{eq:lambda_Di}, we find
	\begin{align}\label{eq:lambdaDi_fix}   \lambda_{\rm Di}\sim\left[0.01\left(\frac{B_{\rm Mpc}}{\rm nG}\right)\right]^{2/(n_{\rm B}+5)}{\rm Mpc}/h.
	\end{align}
	While the above is an order of magnitude estimate, we assume exact equality while establishing the initial conditions. In appendix~\ref{sec:lambda_Di}, we show that $\mathcal{O}(1)$ factor deviation in $\lambda_{\rm Di}$ leads to negligible impact on the power spectrum at lower redshift. Thus, the exact estimation of $\lambda_{\rm Di}$ is likely not important.
	
	In this study, we choose to focus only on two PMF spectral slopes: $n_{\rm B}=2$ and $n_{\rm B}=-2$. In contrast, almost all previous literature that studied PMF's influence on structure formation focused primarily on nearly scale-invariant spectra ($n_{\rm B}\lesssim-3$). We find that our custom version of NGenIC does not accurately produce ICs for nearly scale-invariant spectra, see appendix~\ref{sec:scale_invt}. The $n_{\rm B}=-2$ case is found to be the closest spectra to scale-invariance with marginal numerical issues. Thus we choose $n_B=-2$ as our standard case. Next, we chose to study the $n_{\rm B}=2$ spectrum because it is typically expected in magnetogenesis from phase-transitions \cite{Durrer:2003ja,Subramanian:2015lua,Vachaspati:2020blt}. Furthermore, inverse-cascade in early universe turbulence is also hypothesized to produce spectra of $n_{\rm B}=2$ near the damping scale \cite{Hosking:2022umv}. 
	
	The ICs for baryons and dark matter are given such that the initial density distribution is as smooth as numerically possible, i.e., $\delta \rightarrow 0$. This choice is supported by linear theory, which shows that the post-recombination growth of perturbations is insensitive to initial conditions at recombination (see Appendix~\ref{sec:pre_rec}).

Note that on scales near and smaller than $\lambda_{\rm Di}$, the density perturbations are expected to become non-linear prior to recombination. Here, our linear theory arguments supporting $\delta \rightarrow 0$ in the initial condition are not strictly applicable. Accurately simulating these scales requires simulations that start prior to recombination, which is beyond the current capability of AREPO. Thus, our simulation results are only accurate for scales larger than $\lambda_{\rm Di}$. Readers interested in scales smaller than $\lambda_{\rm Di}$ should instead refer to Refs.~\cite{Jedamzik:2018itu,Jedamzik:2023rfd, Ralegankar:2023pyx}.
	
	\begin{figure}
		\begin{subfigure}{\textwidth}
			\includegraphics[width=1.00\textwidth]{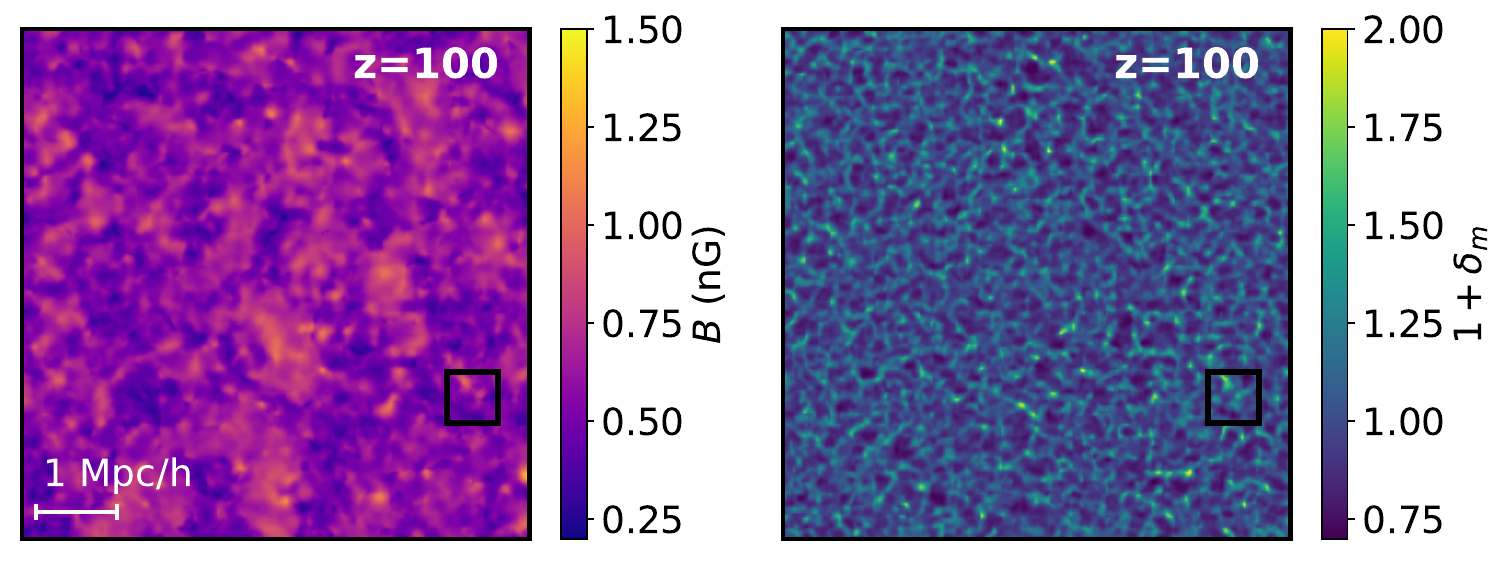}
		\end{subfigure}
		
		\begin{subfigure}{\textwidth}
			\includegraphics[width=1.00\textwidth]{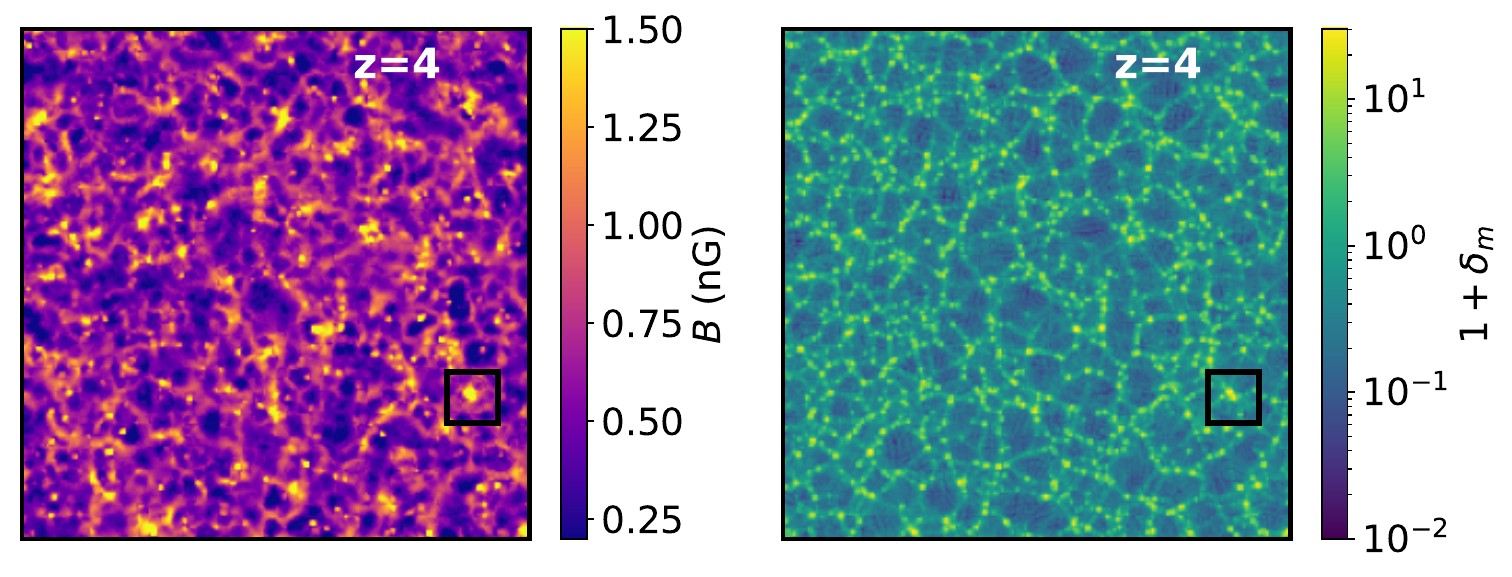}
		\end{subfigure}
		\caption{A thin slice of 0.3 Mpc$/h$ is cut out of the simulation A volume and projected onto a 2D plane. This simulation has $B_{\rm 1Mpc}=0.2$ nG and $n_B=-2$. Left panel shows the map of PMF strength and the right panel shows the overdensity map of the total matter. The limits on the colour-bars are chosen manually for better visualization. These limits are not to be confused with the maximum/minimum values observed in the simulation. The square black box surrounds a region with an overdensity at $z=4$.}\label{fig:slice}
	\end{figure}
	
Additionally, by initially setting $\delta=0$, we are neglecting density perturbations due to inflationary initial conditions. Again, this choice is supported by linear theory, which states that density perturbations sourced by PMFs are insensitive to inflationary initial conditions (see discussion around Eq.~\eqref{eq:lcdm_pmf}). The final matter power spectrum can simply be obtained by summing the power spectrum induced by PMFs with the power spectrum in standard cosmology, as illustrated in Figure 4 of \cite{Ralegankar:2024ekl}. In this study, we are primarily interested in obtaining the matter power spectrum induced by PMFs. In a future study, where we aim to quantify the observable implications of galaxy formation, inflationary initial conditions will also be included in the simulations.

	We choose our box sizes such that the Nyquist scale is just smaller than the viscous damping scale $\lambda_{\rm Di}$. 
	Each simulation box is filled with an (initially-)equal number of gas and DM particles. 
	In table~\ref{table:simparams} we list the different magnetic field configurations and corresponding box sizes for our different simulations. 
	
	\subsection{Non-linear evolution for $n_{\rm B}=-2.0$}\label{sec:nb_minus_2}
\begin{figure}
		\includegraphics[width=1.00\textwidth]{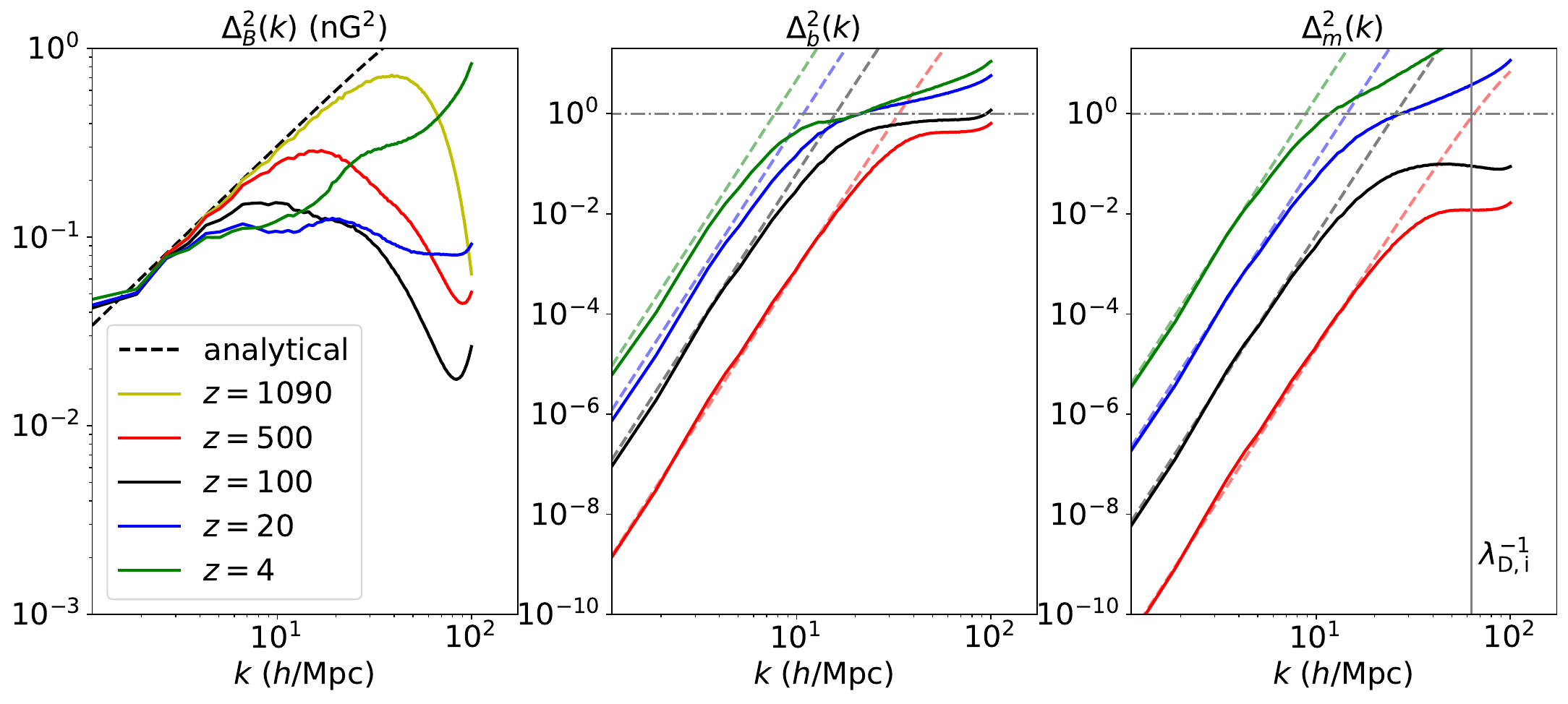}
		\caption{Dimensionless power spectrum at different redshifts for $B_{\rm 1Mpc}=0.2$ nG and $n_{\rm B}=-2$. From left to right we have power spectra of primordial magnetic fields, baryon perturbations, and total matter perturbations. The dashed lines in the panels correspond to the power spectrum computed in the linear limit (eq.~\eqref{eq:Delta_b}). The vertical grey lines mark the scale below which our initial conditions for matter density perturbations are not appropriate.   }\label{fig:nB_minus2_evolve}
	\end{figure}

	In this section, we discuss the non-linear evolution of density perturbation and PMFs using simulation A as a representative. In figure~\ref{fig:slice}, we show the evolution of PMFs and matter perturbations in the simulation. The figure is a zoomed in 2D projection of a thin slice (with depth equal to 0.3 Mpc/$h$) of the whole simulation. In figure~\ref{fig:nB_minus2_evolve}, we show the power spectrum of the magnetic field, baryons, and total matter perturbation at different snapshots for simulation A. 
	
	In the left panel, one can see that the magnetic field power spectrum on small scales is quickly suppressed by non-linear motions in the plasma, while the magnetic fields on large scales remain unaffected throughout the evolution. The suppression on small scales is expected due to MHD turbulence. However, our simulations do not exhibit the expected Kolmogorov power law suppression, $\Delta_{\rm B}\propto k^{-2/3}$. Instead, at $z=500$, we find a steeper suppression with $\Delta_{\rm B}\propto k^{-1}$, and by $z=100$, this suppression becomes shallower, following $\Delta_{\rm B}\propto k^{-1/3}$. Interestingly, for $z<100$, the small-scale ($k>10$ $h/$Mpc) magnetic fields are amplified rather than suppressed.\footnote{The sharp upturn in PMF spectrum at $k\gtrsim 80\, h/$Mpc is a resolution artefact.} This amplification is driven by the collapse of matter under gravity to form halos, where both gravitational compression of baryons and the dynamo mechanism increase the magnetic field energy. This can be further verified by examining the bottom panel of figure~\ref{fig:slice}. Here the large strengths in $B$ fields occur at the same place where there are large overdensities in matter field, which shows that regions of strong magnetic fields correspond to areas of high matter overdensities.
	
	In the centre and right panel of figure ~\ref{fig:nB_minus2_evolve}, we observe that the power spectrum of large-scale density perturbations closely aligns with the linear solutions discussed in the previous section. 
	Notably, this study marks the first time that the analytical linear power spectrum has been successfully verified through MHD simulations.
	
	As expected the power spectrum of baryon perturbations deviates from the linear solution when it reaches non-linear values, $\Delta_{\rm b}\sim \mathcal{O}(1)$. As baryon perturbations are sourcing dark matter perturbations gravitationally, we also see the total matter power spectrum deviating from the linear solution around the same scale where baryon perturbations become non-linear. The effects from baryon thermal pressure are not visible in these simulations because the thermal Jeans scale, $k_{\rm th}\sim 10^3$ 1/Mpc, is not resolved.
	
	Note that $\Delta_{\rm b}$ does not exceed unity prior to $z=100$. This is expected because turbulence in the baryon plasma on small scales should not allow large density fluctuations. However, after $z<100$, the self-gravity of density perturbations overcomes the Lorentz force from PMFs, as discussed in section~\ref{sec:theory}. Consequently, $\Delta_{\rm b}$ can acquire large non-linear values at $z<100$ due to halo formation. 
	
	Halo formation occurs when matter density perturbation exceeds unity. This occurs around
	\begin{align}
		a_{\rm halo}\sim 0.03,
	\end{align}
	or $z_{\rm halo}\sim 30$. The time of halo formation is largely insensitive to both the $B_{\rm 1Mpc}$ as well as $n_{\rm B}$. This insensitivity is because baryon perturbations on small scales are expected to obtain $\sim \mathcal{O}(1)$ values by $z\sim 100$. As gravity always takes over after $z\sim100$, the $\sim \mathcal{O}(1)$ baryon perturbations are found to gravitationally induce $\sim \mathcal{O}(1)$ matter perturbations by $a_{\rm halo}$. If the cosmic baryon fraction was larger, then $z_{\rm halo}$ would have been closer to $z\sim 100$. 
	
	\subsection{Behaviour near the magnetic damping scale}
	As $z=100$ is a special time when gravity overcomes the Lorentz force, we define the magnetic damping scale at that point.
	
	We find the magnetic damping scale, $\lambda_{\rm D}$, by fitting
	\begin{align}\label{eq:PB_powerlaw3}
		P_{\rm B}(k)=Ak^{n_{\rm B}}e^{-k^2\lambda_{\rm D}^2},
	\end{align}
	to the numerical power spectrum at $z=100$. As we are not interested in the detailed turbulent spectrum for $k>\lambda_{\rm D}^{-1}$, the above approximation is suitable for describing magnetic fields on large scales and matches the convention adopted in literature \cite{Planck:2015zrl}. We fit the value of $\lambda_{\rm D}$ by requiring that the total PMF strength in our simulations at $z=100$ matches the PMF strength for the above power spectrum. Or equivalently,
	\begin{align}\label{eq:B_B1mpc}
		\langle B^2\rangle_{z=100}=B_{\rm 1Mpc}^2\left(\frac{\rm Mpc}{\lambda_{\rm D}}\right)^{n_{\rm B}+3}.
	\end{align}
	We find that the above-fitted value of $\lambda_{\rm D}$ agrees with the analytical estimate in eq.~\eqref{eq:lambdaD_va} within 20\%.
	
\begin{figure}
		\includegraphics[width=1.00\textwidth]{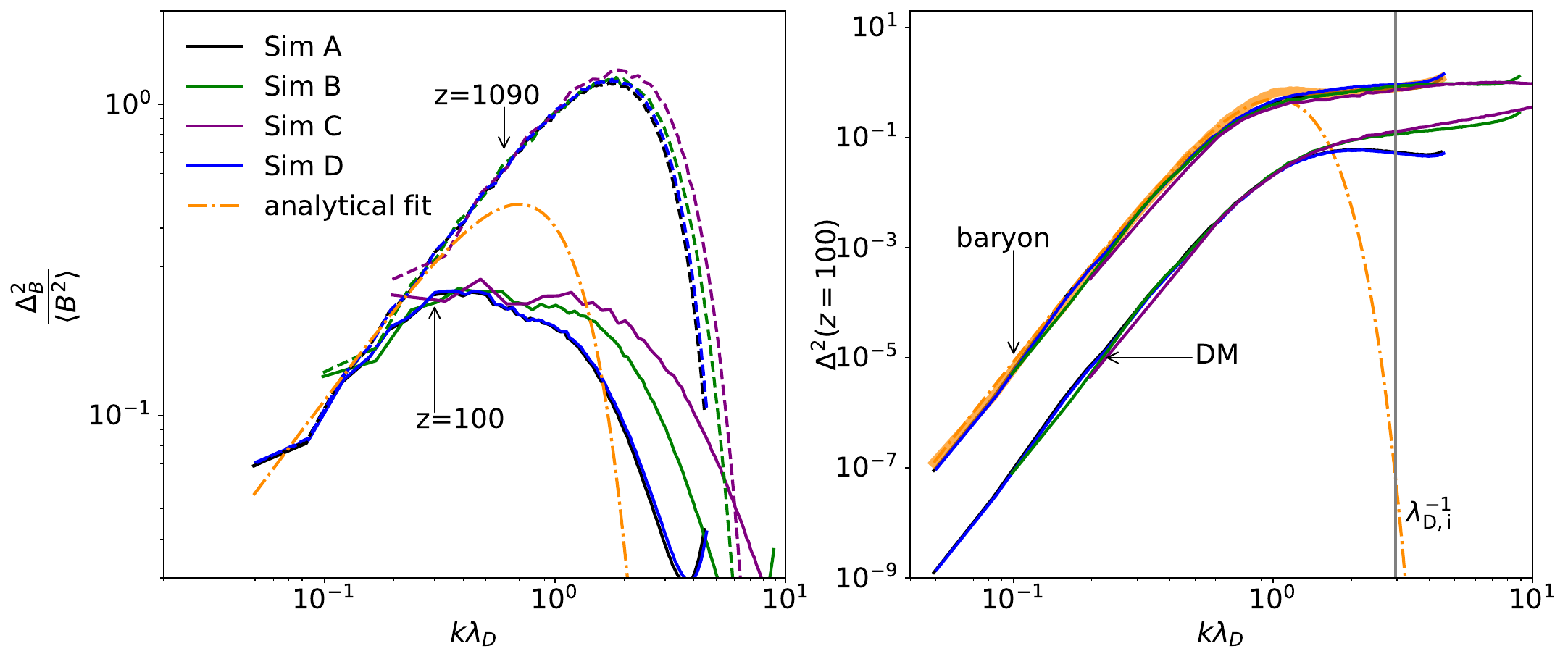}
		\caption{\textbf{Left:} Dimensionless magnetic field power spectrum normalized with total PMF strength, $\langle B^2\rangle$, evaluated at a redshift of 100. The solid lines show spectra at $z=100$, while the dashed lines show spectra at $z=1090$. The $x$-axis has been normalized with $\lambda_{\rm D}$ found using eq.~\eqref{eq:B_B1mpc}. The black, blue, and green lines are for different simulations but all have $n_{\rm B}=-2.0$. As simulation A, B, and C only differ in resolution, we normalize the power spectrum for simulation B and C using the value of $\langle B^2\rangle$ and $\lambda_{\rm D}$ that were found for simulation A. For simulation D, $\langle B^2\rangle$ and $\lambda_{\rm D}$ are evaluated separately. The orange dashed line corresponds to the power spectrum given in eq.~\eqref{eq:PB_powerlaw3} and is matched to have the same $\langle B^2\rangle$ as simulation A. \textbf{Right:} Dimensionless baryon and dark matter power spectrum for different simulations. The colour code is the same as in the left panel. The orange-dashed line is an analytical fit with an exponentially suppressed power on small scales. The thick orange line is the non-linear power spectrum obtained after inputting the analytical power spectrum in NGenIC. The figures highlight that the shape of the power spectra remains unchanged when changing $B_{\rm 1Mpc}$ but changes on small scales with a change in resolution.}\label{fig:compare_sims}
	\end{figure}

	In figure~\ref{fig:compare_sims} we show the magnetic field power spectrum from simulation A as the black solid line in the left panel and compare it with the analytical fitted power spectrum (eq.~\eqref{eq:PB_powerlaw3}) shown as the orange dot-dashed line.
	We find that the turbulent suppression of PMFs starts at scales around three times larger than $\lambda_{\rm D}$. This large deviation from $\lambda_{\rm D}$ is because the orange line has been fitted to yield the correct value of the total PMF strength and not the value of the turbulence scale. As turbulence leads to PMF power being spread almost equally on a wide scale while the fit in eq.~\eqref{eq:PB_powerlaw3} confines power to a much smaller scale, we see the orange line to have a larger peak at a smaller scale.
	
	The scale $\lambda_{\rm D}$ also roughly marks the scale where dark matter and baryon density perturbations deviate from the linear solution, as seen in the right panel of figure~\ref{fig:compare_sims}. 
	Naively, one would have expected the dimensionless baryon power spectrum, $\Delta_{\rm b}$ to be suppressed on scales smaller than $\lambda_{\rm D}$ because turbulence is expected to homogenize density on small scales. However, we instead find $\Delta_{\rm b}$ to saturate to a constant at large $k$.
	
	The value of $\Delta_{\rm b}$ saturates because baryon perturbations are in a mildly non-linear regime at $z=100$. In particular, if we consider the linear baryon power spectrum to have an exponential suppression once $\Delta_{\rm b}$ reaches $\mathcal{O}(1)$ values, then the non-linear power spectrum automatically saturates. This can be seen in the right panel of figure~\ref{fig:compare_sims}, where the orange dot-dashed line is fitted to the baryon power spectrum on large scales through the expression $Cx^ne^{-x^2}$.
	When the fitted function is given as an input power spectrum to NGenIC, the output density fields have a non-linear power spectrum shown by the thick orange line in figure~\ref{fig:compare_sims}. One can see that the thick orange line saturates to a constant at large $k$ where the orange dot-dashed line is suppressed. This shows that the small-scale non-linear power is largely determined by large scales where $\Delta_{\rm b}\sim \mathcal{O}(1)$. 
	
	The reason $\Delta_{\rm m}$ is roughly constant on small scales is because dark matter perturbations are gravitationally sourced by baryon density perturbations. Consequently, the constant value of $\Delta_{\rm b}$ at large $k$ is imprinted onto $\Delta_{\rm DM}$ and hence onto $\Delta_m$. Note that, unlike baryon perturbations, $\Delta_{\rm DM}\ll 1$ at $z=100$ and hence is still in the linear regime.
	
	In other words, there are two effects that lead to deviation from the linear solution: turbulence induced by PMFs and halo formation due to gravity. Turbulence is the primary source of non-linearity at high redshifts ($z>100$) but it only indicates non-linearities in baryons and magnetic fields and not in dark matter perturbations. Whereas halo formation occurs after $z\lesssim 30$ and indicates all perturbations becoming non-linear.
	
	
	\subsection{Dependence on $B_{\rm 1Mpc}$ and resolution}\label{sec:nB_minus2_res}
	Before halo formation, we expect the shape of the power spectra to only be a function of $n_{\rm B}$, whereas changing $B_{\rm 1Mpc}$ is only expected to shift the spectra while keeping its shape unchanged. The horizontal shift can be removed when plotting the spectrum as a function of $k\lambda_{\rm D}$. Moreover, changing the value of $B_{\rm 1Mpc}$ also shifts $\Delta_{\rm B}$ vertically and one can absorb that shift by normalizing the $\Delta_{\rm B}$ with total PMF strength, $\langle B^2\rangle$. This is verified in figure~\ref{fig:compare_sims} where we see simulations A ($B_{\rm 1Mpc}=0.2$ nG) and D ($B_{\rm 1Mpc}=0.8$ nG) to have almost the same shape of the power spectrum for PMFs, baryons, and dark matter.
	
	The insensitivity of the shape of the power spectrum to $B_{\rm 1Mpc}$ implies that the observed peak of the matter power spectrum also would not change with $B_{\rm 1Mpc}$. Consequently, changing $B_{\rm 1Mpc}$ is only expected to change the masses of halos, but their formation time is expected to remain unchanged.
	
	The power spectra at small scales show mild sensitivity to the resolution of the simulations. This is evident in the left panel of figure~\ref{fig:compare_sims}, where simulations B and C use the same values for $B_{\rm 1Mpc}$ and $n_{\rm B}$ as simulation A but with higher resolution. While the large-scale behavior of the PMF spectra is similar in both simulations, at smaller scales where turbulence occurs, we observe greater numerical damping in the lower-resolution simulations.
	
	Because the shape of $\Delta_{\rm B}(z=100)$ differs between simulations A and B, the values of $\langle B^2 \rangle$ and $\lambda_{\rm D}$ are also different. To facilitate comparison, we normalized the spectra of simulation B to match the values of $\langle B^2 \rangle$ and $\lambda_{\rm D}$ found in simulation A.
	
	In contrast to magnetic fields, the baryon power spectrum at $z=100$ is largely insensitive to resolution. This is because the large $k$-value of $\Delta_b$ is determined by the small $k$ where $\Delta_b \sim \mathcal{O}(1)$, meaning increased resolution has little impact on the baryon power spectrum.
	
	Similar to baryon power spectrum, we find that dark matter power spectrum is also not significantly sensitive to resolution. Note that despite simulation A being not well converged, we use it as representative for $n_B=-2$ because it captures a wide range of scales in both the linear and non-linear regimes. Moreover, on scales smaller than $\lambda_{\rm Di}$, the initial conditions for dark matter are inaccurate, meaning that higher-resolution simulations do not provide more accurate results than simulation A. Indeed, Ref.~\cite{Ralegankar:2023pyx} showed that the dark matter power spectrum at scales much smaller than $\lambda_{\rm Di}$ is significantly enhanced compared to larger scales. Accurately quantifying the spectrum at these smaller scales requires tracking the pre-recombination evolution of magnetic fields, which is beyond the scope of this study.
	
	Figure~\ref{fig:compare_sims} also demonstrates that the matter power spectrum is not sensitive to the precise details of MHD turbulence. Despite the slight differences in the turbulent PMF spectra between simulations A and B on small scales, this has no significant effect on the baryon or dark matter power spectrum. Therefore, the fact that our simulations may not fully capture turbulence dynamics should not materially affect our conclusions about the matter power spectrum.
	
	In summary, behaviour on large scales ($l > \lambda_{\rm D}$) is independent of small-scale dynamics and pre-recombination evolution. However, behaviour on intermediate scales ($\lambda_{\rm D} < l < \lambda_{\rm Di}$) shows mild sensitivity to smaller scales and pre-recombination effects. As a result, our findings for scales smaller than $\lambda_{\rm D}$ should be considered as rough estimates, with an uncertainty of approximately $\mathcal{O}(1)$.
	
	%
	
	\subsection{Non-linear evolution for $n_{\rm B}=2$}
	
	\begin{figure}
		\begin{subfigure}{\textwidth}
			\includegraphics[width=1.00\textwidth]{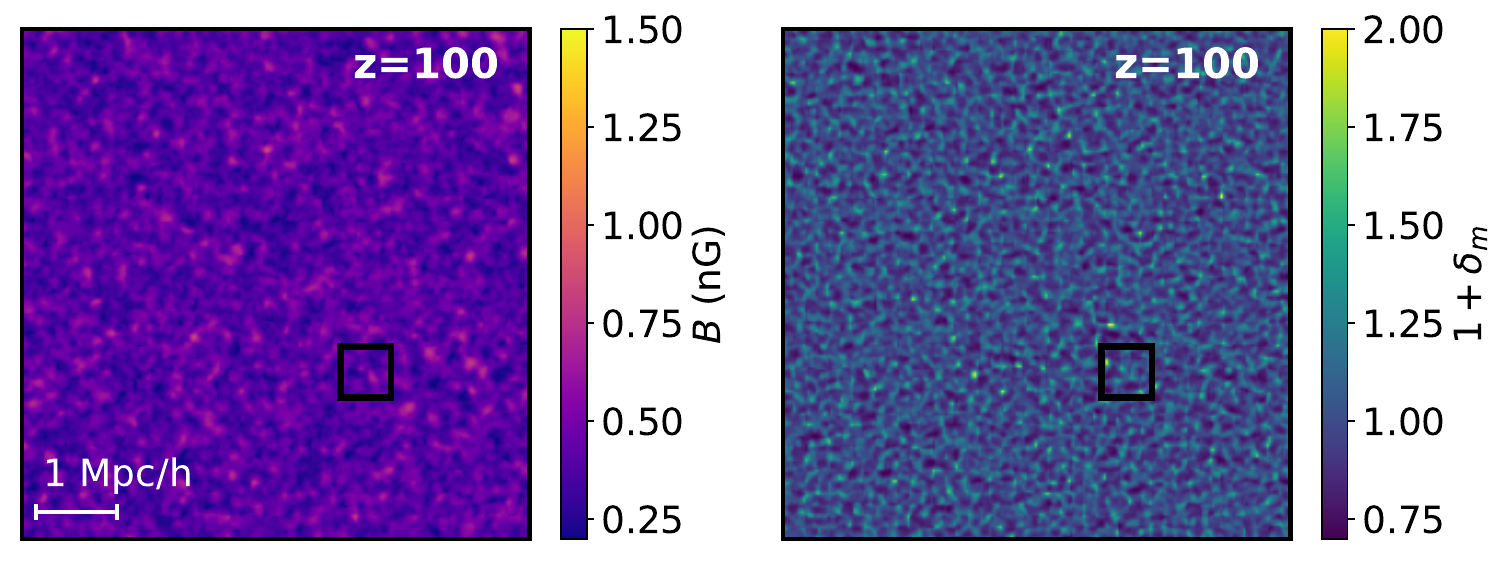}
		\end{subfigure}
		
		\begin{subfigure}{\textwidth}
			\includegraphics[width=1.00\textwidth]{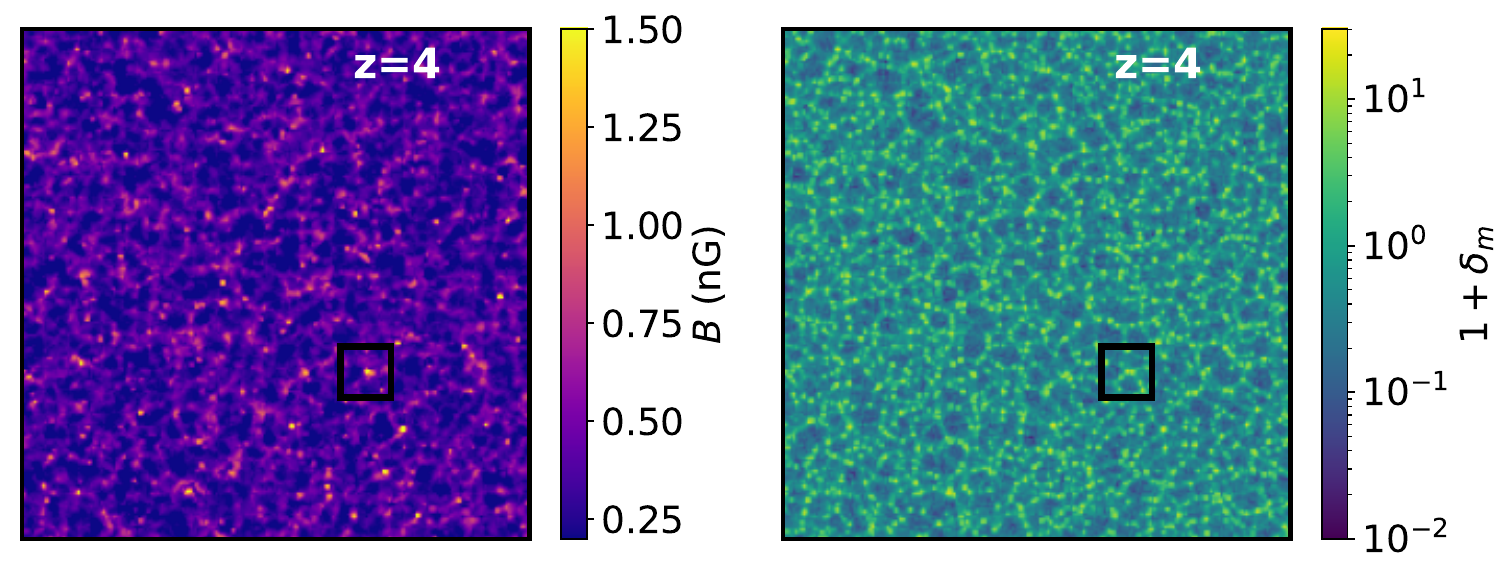}
		\end{subfigure}
		\caption{Same as figure~\ref{fig:slice} but for simulation E ($n_B=2$).}\label{fig:slice_nB2}
	\end{figure}
	
	In section~\ref{sec:theory} we showed that the evolution of density perturbations is qualitatively the same for all $n_{\rm B}$ values much smaller than $-1.5$, where linear theory can be applied on large scales. In the previous sub-section, we focused on $n_{\rm B}=-2$ as a representative spectrum for $n_{\rm B}\ll -1.5$. In this section, we focus on $n_{\rm B}=2$ as a representative spectrum for $n_{\rm B}\gg -1.5$. We choose $n_{\rm B}=2$ also because it is a well-motivated spectrum for PMFs that originate through phase-transitions \cite{Hosking:2022umv,Vachaspati:2020blt,Subramanian:2015lua}. In figure~\ref{fig:slice_nB2}, we show the PMFs and matter perturbations at different redshifts for simulation $E$, which is our representative simulation for $n_B=2$.
	
	\begin{figure}
		\includegraphics[width=1.00\textwidth]{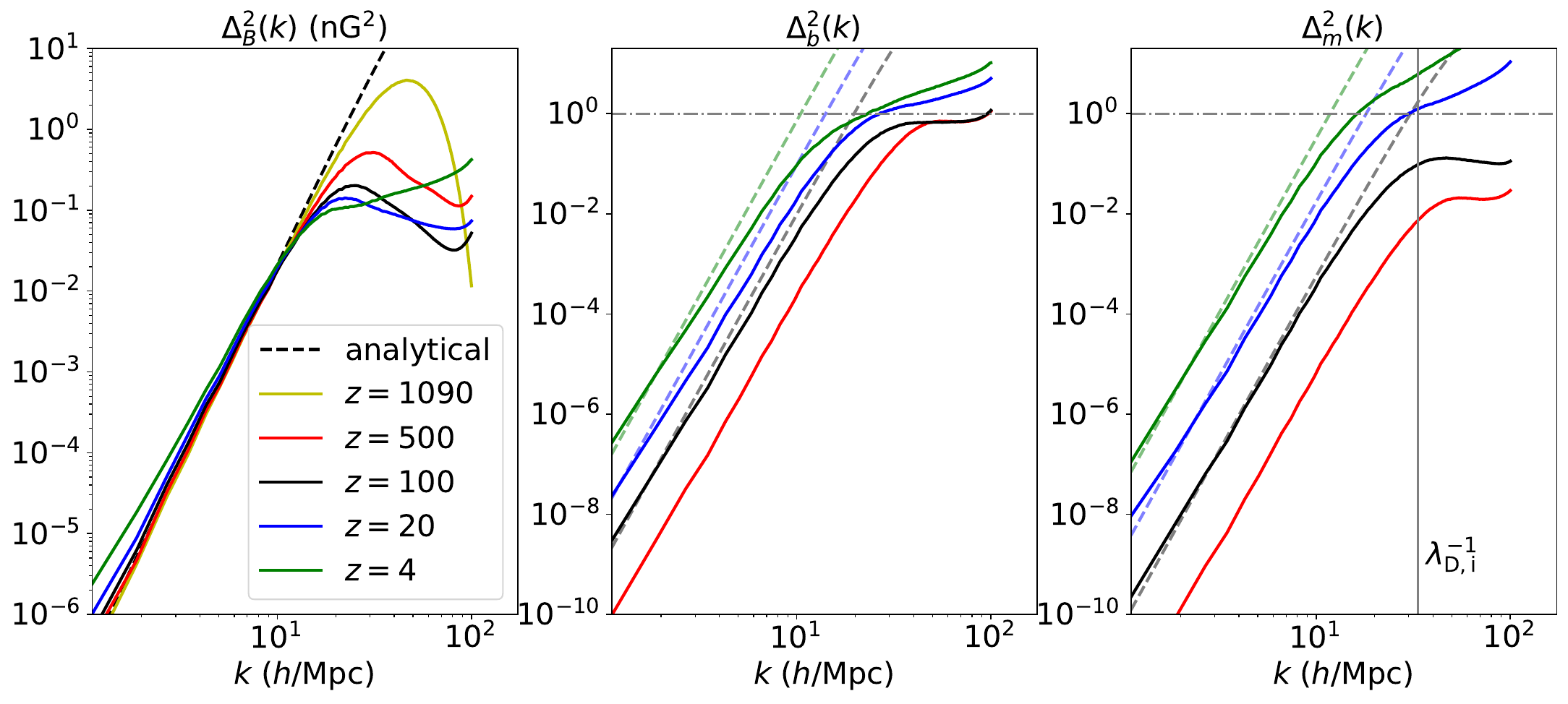}
		\caption{Dimensionless power spectrum at different redshifts for $B_{\rm 1Mpc}=10^{-3}$ nG and $n_{\rm B}=2$ (simulation E in Table~\ref{table:simparams}). From left to right we have power spectra of primordial magnetic fields, baryon perturbations, and total matter perturbations. The dashed line in the left panel shows the PMF power spectrum in linear theory. The dashed lines in the centre and right panels show the fitted function: $c[\xi(z)/\xi(100)]^2k^7$, where the value of $c$ is fitted to the numerical power spectrum at $z=100$.}\label{fig:nB_plus2_evolve}
	\end{figure}
	
	Figure~\ref{fig:nB_plus2_evolve} shows the power spectrum of the magnetic field, baryons, and total matter perturbation at different snapshots for simulation E. The overall qualitative evolution is roughly the same as that found for simulation A in section~\ref{sec:nb_minus_2}: PMFs on small scales are dissipated, while matter perturbations on small scales are enhanced. Furthermore, as in the $n_{\rm B}=-2$ case, we have verified that the shape of the power spectra does not depend on $B_{\rm 1Mpc}$ and the small-scale matter power spectrum is found to be mildly sensitive to resolution. The comparison of the power spectra for simulations with different resolutions is relegated to appendix~\ref{sec:convergence}.
	Now we highlight the key distinctions compared to the $n_{\rm B}=-2$ scenario.
	
	First, we find that the large scale PMFs are not invariant but have a small growth due to inverse cascade. In recent work it was shown that the inverse cascade can be explained through conservation of helicity fluctuations even in a non-helical PMF \cite{Hosking:2022umv}. As our simulations do not accurately resolve the turbulence on small scales, the inverse cascade seen in our simulations may not be quantitatively accurate. 
	
	The magnetic field power spectrum on small scales has much steeper suppression compared to the one observed for $n_{\rm B}=-2$ case. Specifically, we find $\Delta_{\rm B}\propto k^{-5/3}$ on very small scales at $z=100$, which is much steeper than $\Delta_{\rm B}\propto k^{-1}$ observed for $n_{\rm B}=-2$ as well as $\Delta_{\rm B}\propto k^{-2/3}$ expected from Kolmogorov. Again we note that the small-scale turbulent cascade in our simulations may not be accurate because of poor resolution.
	
	Next, the baryon and matter power spectra scale according to the linear theory expectation ($\Delta\propto k^7$) only on very large scales. In figure~\ref{fig:nB_plus2_evolve} the dashed lines in the middle and right panel mark the $k^7$ scaling. One can see that the deviation from $k^7$ scaling occurs on scales about an order of magnitude larger than the turbulence scale. The influence of the turbulence scale extending to such large scales should not be a surprise because even the amplitude of $k^7$ spectrum on large scales is determined by the turbulence scale (see discussion around eq.~\eqref{eq:Pb_int2}). 
	
	To assess whether the growth of matter perturbations aligns with linear theory, we fit the function $ck^7$ individually to both the baryon and total matter power spectra at $z=100$. The fitted function is represented by the black dashed line in Figure~\ref{fig:nB_plus2_evolve}. We then scale this fitted function as $[\xi(a)/\xi(0.01)]^2ck^{7}$ to generate the other coloured dashed lines, where $\xi$ is a solution of eq.~\eqref{eq:deltab_a}.
	Our results indicate that $\xi(a)$ provides a reasonable fit to the time evolution for $z<100$, especially for the total matter power spectrum.

    Finally, unlike the case with $n_B=-2$, the baryon and dark matter power spectra transition into the non-linear regime on wave numbers larger than $1/\lambda_{\rm D,i}$. Our simulation results on these scales are not reliable, see discussion in section~\ref{sec:setup}. However, we do not expect more accurate simulations to qualitatively alter the results on the scales shown in figure~\ref{fig:nB_plus2_evolve}, but exact quantitative values might change.
	
	\section{Semi-analytical fit for the matter power spectrum}\label{sec:fit}
	
	In section~\ref{sec:theory}, we used linear theory to compute the matter power spectrum for any given choice of PMF parameters. While the linear theory allows an easy computation, it fails to provide any estimate for the matter power spectrum in the non-linear regime. Thus, for more accurate behaviour of the matter power spectrum, we resorted to MHD simulations in section~\ref{sec:sim}. Through these simulations, we obtained the matter power spectrum in the non-linear regime for specific chosen PMF parameter points. In this section, we combine the analytical results from section~\ref{sec:theory} with the numerical results from section~\ref{sec:sim} to provide a semi-analytical fit for the baryon and dark matter power spectrum at $z=100$. As gravity dominates over the Lorentz force after $z\lesssim 100$, these semi-analytical fits can be used to provide initial conditions that include the impact of PMFs on the matter power spectrum. Although using only the power spectrum as the initial condition neglects non-Gaussianities and the vortical motion in baryon perturbations, previous work has shown that such initial conditions are sufficient for simulating structure formation \cite{Ralegankar:2024ekl}. Furthermore, one can use the semi-analytical fit to obtain the halo mass function using Sheth-Tormen formalism \cite{Ralegankar:2024ekl}.
	
	We begin by finding an appropriate fitting parameter for the magnetic damping scale $\lambda_{\rm D}$. An order of magnitude estimate for $\lambda_{\rm D}$ can be obtained using eq.~\eqref{eq:lambdaD_va}. We parameterize the unknown coefficient in eq.~\eqref{eq:lambdaD_va} with $\kappa_{\rm D}$, such that
	\begin{align}
		\lambda_{\rm D}=0.1\kappa_{\rm D}(n_{\rm B})\left(\frac{\sqrt{\langle B^2\rangle}}{\rm nG}\right) {\rm  Mpc}.
	\end{align}
	Note that the PMF strength, as well as $\lambda_{\rm D}$, are not exactly constant post-recombination. As gravity is expected to overcome the Lorentz force after $z=100$, we fix the value of $\lambda_{\rm D}$ and $\kappa_{\rm D}$ to be the one at $z=100$, as defined in eq.~\eqref{eq:B_B1mpc}. Replacing $\langle B^2\rangle_{z=100}$ in terms of $\lambda_{\rm D}$ using eq.~\eqref{eq:B_B1mpc}, we obtain
	\begin{align}\label{eq:lambdaD_fix}   
		\lambda_{\rm D}=\left[0.1\kappa_{\rm D}(n_{\rm B})\left(\frac{B_{\rm Mpc}}{\rm nG}\right)\right]^{2/(n_{\rm B}+5)}{\rm Mpc}.
	\end{align}
	Matching the above expression of $\lambda_{\rm D}$ to the $\lambda_{\rm D}$ values in simulation A and E, we find
	\begin{align}
		\kappa_{\rm D}(n_{\rm B}=-2)=0.8 && \kappa_{\rm D}(n_{\rm B}=2)=0.9.
	\end{align}
	Note that $\kappa_{\rm D}$ is not dependent on $B_{\rm 1Mpc}$ because $\kappa_{\rm D}$ is essentially a parameter for the shape of the PMF spectrum near $\lambda_{\rm D}$ and $B_{\rm 1Mpc}$ does not control the shape of the spectrum. 
	
	
	Next, we provide fitting functions for the dimensionless power spectra of matter fields. For $n_{\rm B}<-1.5$, matter fields on large scales are well described by the linear spectrum given in section~\ref{sec:theory} and are proportional to $k^{2n_{\rm B}+10}$. On small scales, the power spectra seem to have a proportionality of $\propto k^0$. Consequently, we fit the power spectrum through
	\begin{align}\label{eq:delta_dmfit}
		\Delta^{\rm fit}=\Delta^{\rm lin}(k,a)\left[1+(k\lambda_{\rm J})^{p}\right]^{-(2n_{\rm B}+10)/p},
	\end{align}
	where $\Delta^{\rm lin}$ is determined by eq.~\eqref{eq:Delta_b}.
	In the above fitting function, $\lambda_{\rm J}$, parameterizes the scale where the power spectrum deviates from the linear solution and $p$ parameterizes the sharpness of the deviation. We further re-parameterize $\lambda_{\rm J}$ similar to $\lambda_{\rm D}$ with
	\begin{align}\label{eq:lambdaJ_fix}
		\lambda_{\rm J}=\left[0.1\kappa_{J}(n_{\rm B})\left(\frac{B_{\rm 1Mpc}}{\rm nG}\right)\right]^{2/(n_{\rm B}+5)}{\rm Mpc}.
	\end{align}
	The above parameterization allows us to have the fitting parameters, $\kappa_{\rm J}$ and $p$, independent of $B_{\rm 1Mpc}$.
	
	\begin{figure}
		\begin{subfigure}{1.00\textwidth}
			\includegraphics[width=1.00\textwidth]{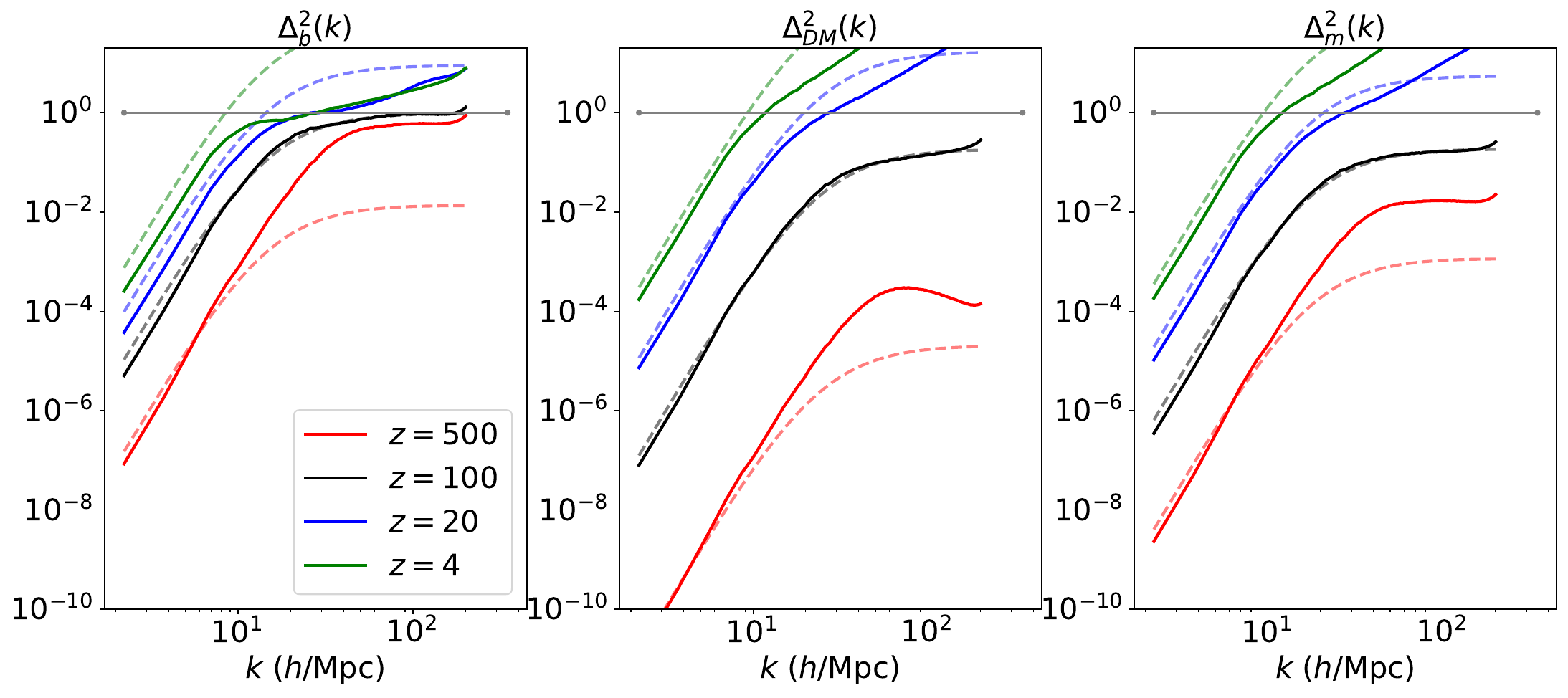}
		\end{subfigure}
		
		\begin{subfigure}{1.00\textwidth}
			\includegraphics[width=1.00\textwidth]{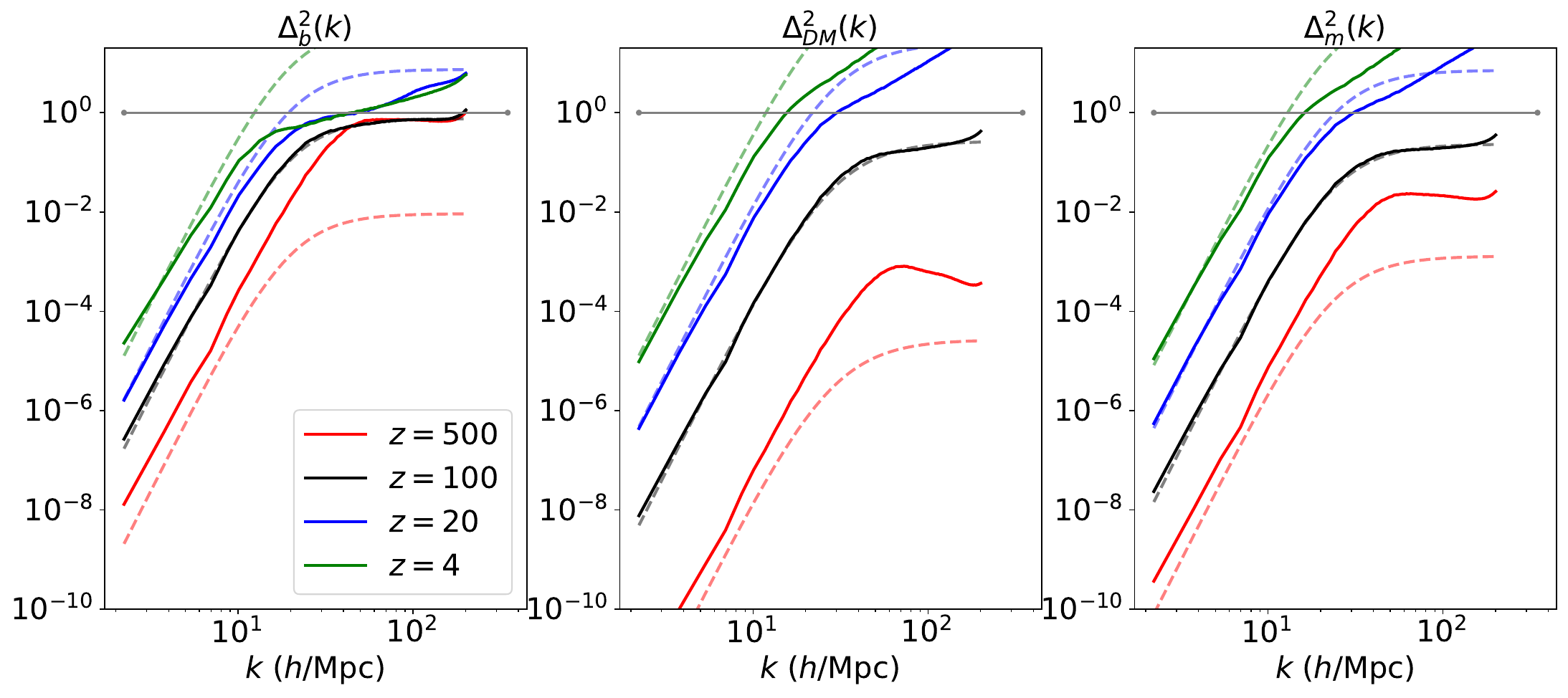}
		\end{subfigure}
		\caption{Comparison of the analytically fitted dimensionless power spectrum to the numerical one for simulation B (top) and simulation F (bottom). The solid lines are from our simulations while the dashed lines are the analytical fits obtained using eqs.~\eqref{eq:delta_dmfit}-\eqref{eq:gamma_fit}.}\label{fig:nB_fit}
	\end{figure}
	
	We fit the values of $\kappa_{\rm J}$ and $p$ to the numerically obtained power spectrum at $z=100$. 
	For $n_{\rm B}=-2$, we use simulation B to find the fitting parameters and obtain
	\begin{align}
		\kappa_{\rm J}^{b}=1.57 && \kappa_{\rm J}^{\rm DM}= 0.774&& \kappa_{\rm J}^{m}=1.17\\
		p_b=1.95 && p_{\rm DM}=1.86 && p_m=1.94.
	\end{align}
	Note that we use simulation B instead of A because the matter power spectrum of A is not well converged on small scales.
	
	We show in the top panel of figure~\ref{fig:nB_fit} how the analytical fit compares with the numerical power spectrum. The extrapolation of the fit at $z=100$ to smaller redshifts using $\xi(a)$ is not precise when $\Delta>1$. This is because our semianalytical fits are primarily made to capture the effect of turbulence at high redshifts ($z\sim100$) and not the non-linearities induced by halo formation. However, the extrapolated fits can be used to obtain an approximate halo mass function using Sheth-Tormen formalism.
	
	Note that for $n_{\rm B}>-1.5$ the fitting function in eq.~\eqref{eq:delta_dmfit} is not appropriate as even the large scale matter power spectrum is determined by PMF spectra on $\lambda_{\rm D}$ scales. Thus, apart from $p$ and $\lambda_{\rm J}$, we require an additional fitting parameter to quantify the power spectrum on large scales. Specifically, we fit the power spectrum for matter fields at $z=100$ using
	\begin{align}\label{eq:fit_nB_2}
		\Delta^{\rm fit}= \gamma \frac{(k\lambda_{\rm J})^7}{[1+(k\lambda_{\rm J})^p]^{7/p}}\ ,
	\end{align}
	where $p$, $\gamma$, and $\lambda_{\rm J}$ are the parameters to be fitted. For $n_{\rm B}=2$ we use simulation F for our fitting parameters and find
	\begin{align}
		\kappa_{\rm J}^{b}=1.13 && \kappa_{\rm J}^{\rm DM}=0.323 && \kappa_{\rm J}^{m}=0.593  \\
		p_b=2.50 && p_{\rm DM}=2.17 && p_m=2.36\\
		\gamma_b=0.787 && \gamma_{\rm DM}=0.286 && \gamma_m=0.244.\label{eq:gamma_fit}
	\end{align}
	We extrapolate the above power spectrum to other redshifts using $\Delta^{\rm fit}\times [\xi(a)/\xi(0.01)]^2$, where $\xi(a)$ is the function derived using linear theory (see section~\ref{sec:theory}). A better extrapolation can be obtained by solving the differential equations ~\eqref{eq:deltab_a} and \eqref{eq:deltadm_a} with $\Delta^{\rm fit}$ as the initial condition at $z=100$.  
	In the bottom panel of figure~\ref{fig:nB_fit}, by comparing the solid and dashed curves, it is possible to appreciate how the analytical fit performs with respect to the numerical power spectrum.
	
	The value of these fitted parameters only depends on $n_{\rm B}$ and not on $B_{\rm 1Mpc}$. However, the fitted parameters are affected by the values of the fields near the $\lambda_{\rm Di}$ scale, which are not precisely known. Moreover, the resolution of the simulations is not enough to analyze the power spectrum around $\lambda_{\rm Di}$. Thus, the power spectrum obtained using the above-fitted parameters should only be taken as $\mathcal{O}(1)$ estimate of the actual matter power spectrum on non-linear scales.
	
	In the previous literature, it was expected that the matter power spectrum is suppressed roughly below the scale $\lambda_{\rm J}$ \cite{Kim:1994zh, Sethi:2004pe, Cruz:2023rmo, Shaw:2010ea, Ralegankar:2024ekl}. Specifically, these studies drew an analogy with the thermal Jeans scale and claimed that the growth of density perturbations should be suppressed on scales where the magnetic pressure overcomes gravity. For instance, by replacing the baryon sound speed with the Alfvén speed in the thermal Jeans scale, Ref.~\cite{Kim:1994zh} found the magnetic Jeans scale to be
	\begin{align}\label{eq:naive_jeans}
		\lambda_{\rm J}\sim\frac{v_{\rm A}}{a}\sqrt{\frac{\pi}{G\rho_{\rm m}}}=2\pi\sqrt{\frac{2}{3}}\frac{v_{\rm A}}{aH}.
	\end{align}
	In the last line, we used the fact that in a matter-dominated universe, $H^2=8\pi G\rho_{\rm m}/3$. Other studies use similar arguments to calculate the magnetic Jean scale, albeit with different $\mathcal{O}(1)$ factors \cite{Sethi:2004pe, Cruz:2023rmo, Shaw:2010ea}. The above expression of the magnetic Jeans scale can be rewritten in the form of eq.~\eqref{eq:lambdaJ_fix} but with different values of $\kappa_{\rm J}$.
	
	The precise estimate of $\kappa_{\rm J}$ is important because the matter power spectrum on non-linear scales is highly sensitive to its value. Considering $n_{\rm B}<-1.5$, and approximating the power spectrum to be equal to the value of the linear solution at $k=\lambda_{\rm J}^{-1}$, one finds
	\begin{align}
		\Delta^{\rm lin}(k=\lambda_{\rm J}^{-1})\propto \kappa_{\rm J}^{-4}.
	\end{align}
	Thus, even a factor of 2 deviation in $\kappa_{\rm J}$ can introduce an order of magnitude deviation in $\Delta$ in the non-linear regime.
	
\begin{figure}
		\includegraphics[width=1.00\textwidth]{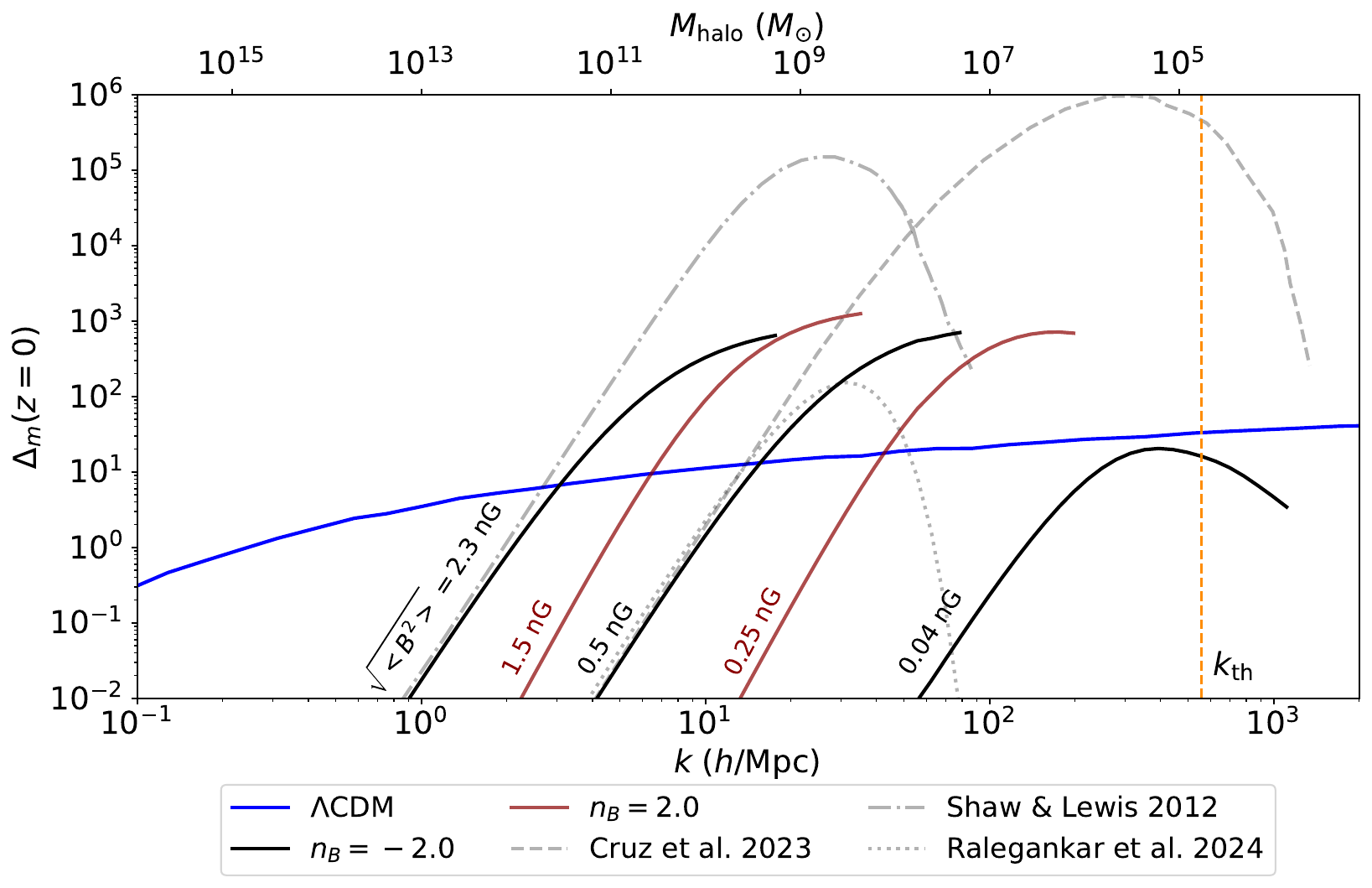}--
		\caption{Dimensionless total matter power spectrum extrapolated to $z=0$. The shown power spectra capture the non-linearity due to baryons back-reacting on the magnetic fields but not the non-linearity due to halo formation. The solid black and brown lines show the spectra derived using analytical fits discussed in section~\ref{sec:fit}. From left to right the corresponding value of $B_{\rm 1Mpc}$ is 1 nG, $10^{-3}$ nG, 0.1 nG, $2\times 10^{-5}$ nG, and $2\times 10^{-3}$ nG. In the plot, we also report the value of total comoving magnetic field strength at $z=100$. The blue curve is the $\Lambda$CDM prediction. The grey lines are power spectra obtained from earlier literature: dot-dashed from Ref.~\cite{Shaw:2010ea}, dashed from Ref.~\cite{Cruz:2023rmo}, and dotted from Ref.~\cite{Ralegankar:2024ekl}. The vertical orange dashed line is the baryon thermal Jeans scale. One can see that previous studies miscalculated the peak of the matter power spectrum by orders of magnitude.}\label{fig:Pk_final} 
	\end{figure}

	In figure~\ref{fig:Pk_final} we show the matter power spectrum extrapolated to $z=0$ from earlier literature as grey lines.\footnote{Specifically, we obtain the dashed and dot-dashed lines from figure~2 of Ref.~\cite{Cruz:2023rmo} and figure~3 of Ref.~\cite{Shaw:2010ea}, respectively. As Ref.~\cite{Shaw:2010ea} does not provide the power spectrum for $n_B=-2$, we use their $n_B=-2.1$ power spectrum when plotting the dot-dashed gray line. Moreover, Ref.~\cite{Shaw:2010ea} follow a slightly different convention for the definition of $B_{\rm 1Mpc}$. We find that our matter power spectrum agrees with their result for $B_{\rm 1Mpc}$ values two times smaller than their $B_{\rm 1Mpc}$. The dotted gray line has been obtained using eq.~(2.23) of Ref.~\cite{Ralegankar:2024ekl}.} We compare these power spectra to the semi-analytically computed power spectra (solid lines). One can see that several previous works overestimated the matter power spectrum on small scales by several orders of magnitude. Thus, it is likely that constraints on PMFs derived from Lyman-$\alpha$, reionization, or abundance of dwarf galaxies are also over-estimated and need to be recomputed.
	
	Note that while the total matter power spectrum is useful for obtaining the halo mass function, it cannot provide an estimate of the baryon fraction in the halos \cite{Ralegankar:2024ekl}. Therefore, in future studies that aim to utilize our fitting functions, we emphasize that both the baryon and dark matter power spectra should be used if the final observable is sensitive to baryons.
	
	We only show the semi-analytical power spectrum for $k$ values $k\lesssim \lambda^{-1}_{\rm D}$.
	For $k\gg \lambda^{-1}_{\rm D}$, the matter power spectrum is further significantly amplified once we take into account pre-recombination evolution \cite{Ralegankar:2023pyx}. As this study is focused on post-recombination evolution, we chose to consider only Fourier modes with $k\lesssim \lambda^{-1}_{\rm D}$.
	
	Additionally, the semi-analytical fits derived in this section are applicable as long as $\lambda_{\rm D}$ is much larger than the thermal Jeans scale, $\lambda_{\rm th}$. For $\lambda_{\rm D}<\lambda_{\rm th}$, the turbulence induced by PMFs occurs in the incompressible regime. In contrast, our MHD simulations were primarily focused on compressible fluid. Consequently, the values of our fitted parameters could change for $\lambda_{\rm D}<\lambda_{\rm th}$. However, since the matter power spectrum induced by PMFs is suppressed below the $\Lambda$CDM power spectrum in this regime, precise computation of parameters for $\lambda_{\rm D} < \lambda_{\rm th}$ is not crucial.
	
	\section{Summary and Conclusions}\label{sec:con}
	This study presents the first-ever (to the best of our knowledge) coupled gravity and MHD simulations of primordial magnetic fields from recombination time, as well as their time-dependent impact on the matter power spectrum. We focus only on perturbations induced by PMFs and neglect the nearly scale-invariant $\Lambda$CDM perturbations. Furthermore, we neglect the pre-recombination evolution of perturbations, which, we argued using linear theory, is only expected to introduce $\mathcal{O}(1)$ correction on scales of our interest.
	
	We identify three distinct phases in the post-recombination evolution ($z<$1100) for PMFs with strengths larger than $\sim0.05$ nG. In the first phase, PMFs induce turbulent motion in the plasma, leading to baryon density perturbations reaching $\mathcal{O}(1)$ values on scales smaller than the turbulence scale and a power-law suppression on larger scales. During this phase, PMFs evolve under a turbulent cascade. The first phase ends around $z\sim 100$ when gravity overcomes the Lorentz force and the turbulent evolution is reduced. In this second phase, density perturbations on all scales grow under gravity while the spectrum of magnetic fields remains roughly unchanged. Gravity eventually causes the total matter perturbations to collapse and form halos by $z \sim 30$. In this final phase ($z\gtrsim 30$), halo formation leads to the growth of magnetic fields by the dynamo mechanism. The timing of these phases is largely independent of the strength and spectrum of the PMFs, with smaller PMFs shifting the turbulence to smaller scales. PMFs weaker than $\sim$0.05 nG induce turbulence on scales smaller than the baryon thermal Jeans scale and hence fail to generate significant density perturbations.
	
	Beyond providing a clear narrative of matter evolution under PMFs, our study also delivers several new quantitative results. First, our MHD simulations confirm the predictions of linear theory for large-scale perturbations induced by PMFs.
	
	Next, our simulations explore the power spectrum for magnetic fields with a Batchelor spectrum, which is commonly associated with magnetogenesis scenarios such as phase transitions. This analysis, which cannot be captured by linear theory even on large scales, significantly broadens the parameter space for PMFs.
	
	One of the key findings of this work lies in the behaviour of the matter power spectrum in the non-linear regime. Contrary to earlier studies, we found that the baryon power spectrum does not exhibit suppression at scales smaller than the magnetic Jeans scale. Additionally, previous studies were found to have overestimated the total matter power spectrum near the magnetic Jeans scale by orders of magnitude. This likely led to an overestimation of the constraints on PMFs coming from structure formation (Lyman-$\alpha$ forest, reionization, galaxy abundance, etc.), suggesting that these constraints may need to be revisited.
	
	To facilitate a better evaluation of future constraints on PMFs, we provide fitting functions for the baryon, dark matter, and total matter power spectra. These functions can be used as initial conditions for simulations that do 
	not have MHD but aim to explore the impact of the enhanced matter power spectrum caused by PMFs. However, we stress that results from such simulations should only be used as a guideline, as many astrophysical signals could be influenced by magnetic fields. Moreover, the fitting functions we provide should only be taken as $\mathcal{O}(1)$ estimate of the power spectrum in the non-linear regime. Detailed numerical simulations that track the transition through the epoch of recombination would be required to precisely estimate the power spectrum on these small scales.
	
	
	The primary goal of this work is to provide an estimate of the matter power spectrum induced by PMFs through comprehensive state-of-the-art numerical simulations. A detailed exploration of how this matter power spectrum impacts cosmological observables will be addressed in future work.
	
	\appendix
    \section{Pre-recombination evolution in linear limit}\label{sec:pre_rec}
    In this appendix, we review the evolution of perturbations in the pre-recombination universe and show that the pre-recombination evolution is largely unimportant for the scales of our interest. 

   Before recombination, baryons are tightly coupled to photons via Thomson scattering, creating two distinct regimes based on the photon mean free path:
    \begin{align}
        l_{\gamma}=\frac{1}{n_e\sigma_Ta}.
    \end{align}
    For scales larger than $l_{\gamma}$, baryons and photons behave as a single fluid, where relativistic thermal pressure prevents the growth of PMF-induced perturbations. 
    On scales smaller than $l_{\gamma}$, the baryon perturbations are effectively decoupled from photon perturbations and hence the baryon fluid no longer has relativistic thermal pressure. Consequently, PMFs can induce significant density perturbations on scales below the photon mean free path.

    We shall focus on the evolution of baryon perturbations below the photon mean free path and under the influence of PMFs. On these scales, the free streaming of photons damps out photon perturbations, creating an effectively homogeneous photon background for baryons. The Thompson scattering of baryons with the photon background leads to a viscous drag for baryons. Thus, the baryon Euler equation given in eq.~\eqref{eq:thetab} is modified to \cite{Subramanian:1997gi}
    \begin{align}\label{eq:thetab2}
		\frac{\partial \vec{v}_{\rm b}}{\partial t}+H\vec{v}_{\rm b}+\alpha\vec{v}_b+\frac{(\vec{v}_{\rm b}\cdot\nabla)\vec{v}_{\rm b}}{a}+\frac{c_{\rm b}^2}{a}\nabla \delta_{\rm b}&=\frac{(\nabla\times\vec{B})\times\vec{B}}{4\pi a^5\rho_{\rm b}}-\frac{\nabla\phi}{a},
    \end{align}
    where
    \begin{align}
        \alpha=\frac{4\rho_{\gamma}}{3\rho_bl_{\gamma}}
    \end{align}
    is the photon drag term. All the other relevant MHD equations (eqs.~\eqref{eq:induction}-\eqref{eq:contb}) are unaffected by photon drag. Before recombination, $\alpha$ is significantly larger than $H$ with $\alpha/H\sim 350 (a_{\rm rec}/a)^2$. At recombination, the electron fraction drops significantly, which almost instantaneously reduces $\alpha$ below the Hubble rate. Thus the photon drag is almost instantaneously removed after recombination.
    

Considering the linear limit, we obtain the same governing equations as those given in eqs.~\eqref{eq:constB}-\eqref{eq:deltadm_a}, except the equation for baryon is modified to
    \begin{align}\label{eq:deltab_a2}
		a^2\frac{\partial^2 \delta^{\rm lin}_{\rm b}}{\partial a^2}+a\left[\frac{3}{2}+\frac{\alpha}{H}\right]\frac{\partial \delta^{\rm lin}_{\rm b}}{\partial a}+\frac{c_{\rm b}^2}{(aH)^2}k^2 \delta^{\rm lin}_{\rm b}-\frac{3}{2}\frac{\Omega_{\rm b}}{\Omega(a)}\delta^{\rm lin}_{\rm b}=-\left[\frac{3M_{\rm Pl}^2S_B}{\rho_{\rm m0}}\right]\frac{\Omega_{\rm m}}{\Omega(a)}+\frac{3}{2}\frac{\Omega_{\rm DM}}{\Omega(a)}\delta^{\rm lin}_{\rm DM}.
	\end{align}

    Note that the above equation is only valid below the photon mean free path. Let $a_{\rm mfp}$ be the scale factor when 
    $$l_{\gamma}(a_{\rm mfp})=k^{-1}.$$ 
    Then eq.~\eqref{eq:deltab_a2} and eq.~\eqref{eq:deltadm_a} are to be solved only for $a>a_{\rm mfp}$. As photon thermal pressure inhibits PMFs from sourcing inhomogeneitis for $a<a_{\rm mfp}$, we can use the initial condition where
    \begin{align}\label{eq:ics2}
		\delta^{\rm lin}(a_{\rm mfp})=0 && \frac{\partial\delta^{\rm lin}(a_{\rm mfp})}{\partial a}=0,
	\end{align}
    for both baryons and dark matter. Note that we only focus on density perturbations sourced by PMFs ($S_B$ term) and ignore the contribution from inflationary initial conditions. We can do so because, for ordinary differential equations, the homogeneous solution sourced by initial conditions has no influence on the inhomogeneous solution imposed by an external source ($S_B$).

	\begin{figure}
        \centering
			\includegraphics[width=\textwidth]{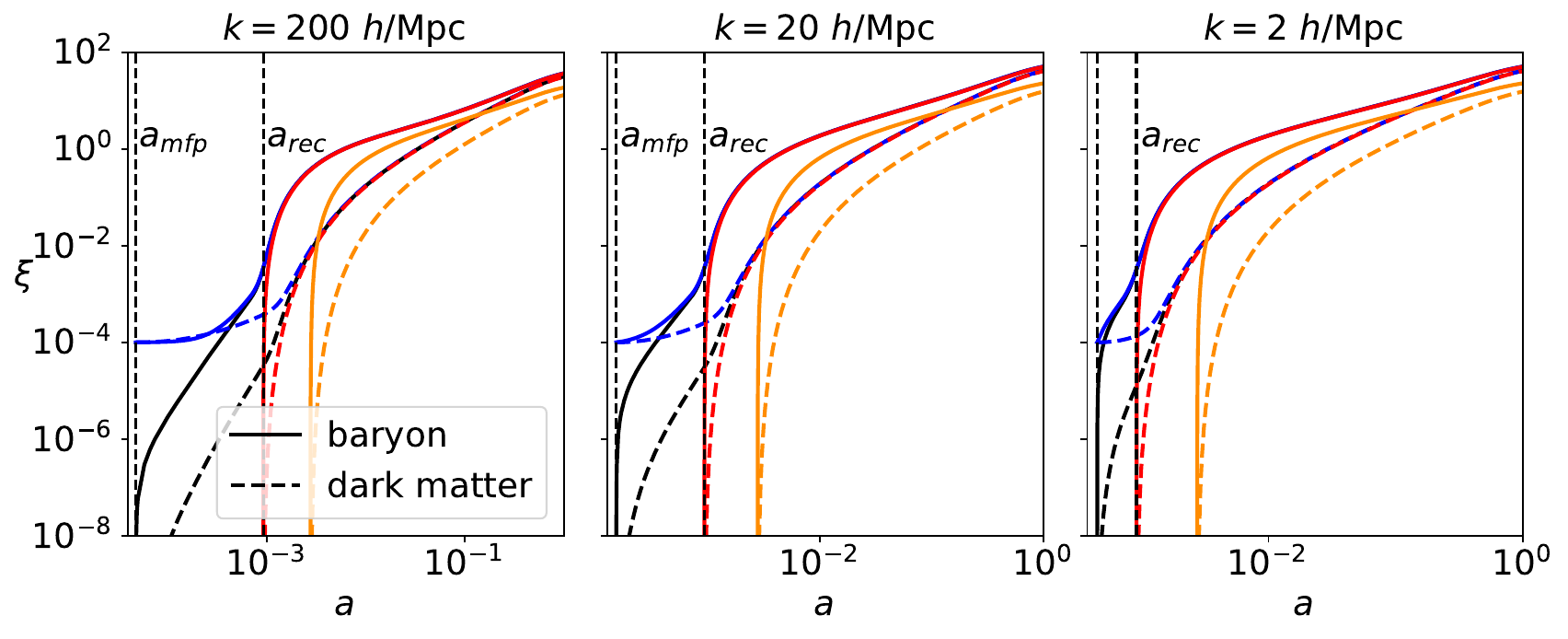}
		\caption{\textbf{Left}: Evolution of baryon (solid) and dark matter (dashed) density perturbations normalised with $3M_{\rm Pl}^2S_B/[a^3\rho_{\rm m}]$, which parameterises the Lorentz force. Different colours indicate solutions with different initial conditions. Black lines consider $\delta(a_{\rm mfp})=0$ for both dark matter and baryons. Similarly, blue lines consider $\delta(a_{\rm mfp})=10^{-4}$, red lines consider $\delta(a_{\rm rec})=0$, and orange lines consider $\delta(3a_{\rm rec})=0$. The perturbations are for Fourier modes with wave number $k=200\ h$ Mpc${}^{-1}$ (left), $k=20\ h$ Mpc${}^{-1}$ (center), and $k=2\ h$ Mpc${}^{-1}$ (right). One can see that the final value of $\xi$ is insensitive to the exact choice of initial conditions prior to recombination but is sensitive to initial conditions after recombination.}\label{fig:xi_pre_rec}
	\end{figure}

    By replacing $\delta$ in terms of $\xi$,
    \begin{align}
		\delta_{\rm b}^{\rm lin}=-\xi_{\rm b}(k,a)\frac{3M_{\rm Pl}^2}{\rho_{\rm m0}}S_B && \delta_{\rm DM}^{\rm lin}=-\xi_{\rm DM}(k,a)\frac{3M_{\rm Pl}^2}{\rho_{\rm m0}}S_B,
	\end{align}
    we can directly solve  eq.~\eqref{eq:deltab_a2} and eq.~\eqref{eq:deltadm_a} once the value of $k$ is specified. In figure~\ref{fig:xi_pre_rec} we show the evolution $\xi$ starting from $a_{\rm mfp}$. The sharp increase in $\xi_b$ near recombination is caused due to the sharp decrease in $\alpha$ at recombination.

    In figure~\ref{fig:xi_pre_rec} we also show the solution to $\xi$ for different initial conditions. One can see that the final solution is insensitive to the initial conditions prior to recombination. However, different initial conditions post-recombination have different final solution. Thus, the concerned equations have an attractor solution before recombination. This validates our choice of the initial conditions used in the main body (eq.~\eqref{eq:delta_ic}).

    The argument for the attractor solution is appropriate as long as perturbations are small and in the linear regime. Initially, near $a_{\rm mfp}$ density perturbations are expected to be very small due to silk damping. Thus, the initial condition in eq.~\eqref{eq:ics2} is always appropriate. In contrast, the initial condition given in eq.~\eqref{eq:delta_ic} is only appropriate if $\delta_b$ remains in the linear regime prior to recombination. That is we require, 
    \begin{align}
        \delta_b(a_{rec})=\xi_b(a_{rec})\frac{3M_{\rm Pl}^2}{\rho_{m0}}S_B\ll1.
    \end{align}
    Using the fact that $\xi_b(a_{rec})\sim 10^{-3}$ and that
    \begin{align}
        S_B\sim l^2\frac{\sqrt{\langle B^2\rangle}}{4\pi\rho_{b0}},
    \end{align}
    where $l$ is the characteristic length scale, $\delta_b(a_{rec})\ll 1$ can be rewritten as
    \begin{align}
        l^2\gg 10^{-3}\left(\frac{\sqrt{\langle B^2\rangle}}{4\pi\rho_b}\right)\left(\frac{3M_{\rm Pl}^2}{a^2\rho_{m}}\right). 
    \end{align} 
   Using present-day values of $\rho_{\rm m0}\approx 1.15\times 10^{-47} {\rm GeV}^4$ and $\rho_{\rm b0}\approx 1.64\times10^{-48} {\rm GeV}^4$, the above becomes
    \begin{align}
        l\gg 0.01 {\rm Mpc} \left(\frac{\sqrt{\langle B^2\rangle}}{\rm nG}\right)=\lambda_{\rm D,i}.
    \end{align}
    The quantity on the right is the same as the magnetic damping scale just before recombination \cite{Jedamzik:1996wp, Subramanian:1997gi}. This is not a coincidence: the magnetic damping scale reflects the length scale where baryonic perturbations become non-linear and begin to back-react on the magnetic fields.

The sharp increase in $\xi_b$ near recombination also causes a significant rise in the magnetic damping scale. This explains the order-of-magnitude difference between the pre-recombination damping scale ($\lambda_{\rm D,i}$) and the post-recombination scale ($\lambda_{\rm D}$) as seen in eqs.~\eqref{eq:lambdaD_va} and \eqref{eq:lambda_Di}.

	\section{Dependence on $\lambda_{\rm Di}$}\label{sec:lambda_Di}
\begin{figure}
		\includegraphics[width=1.00\textwidth]{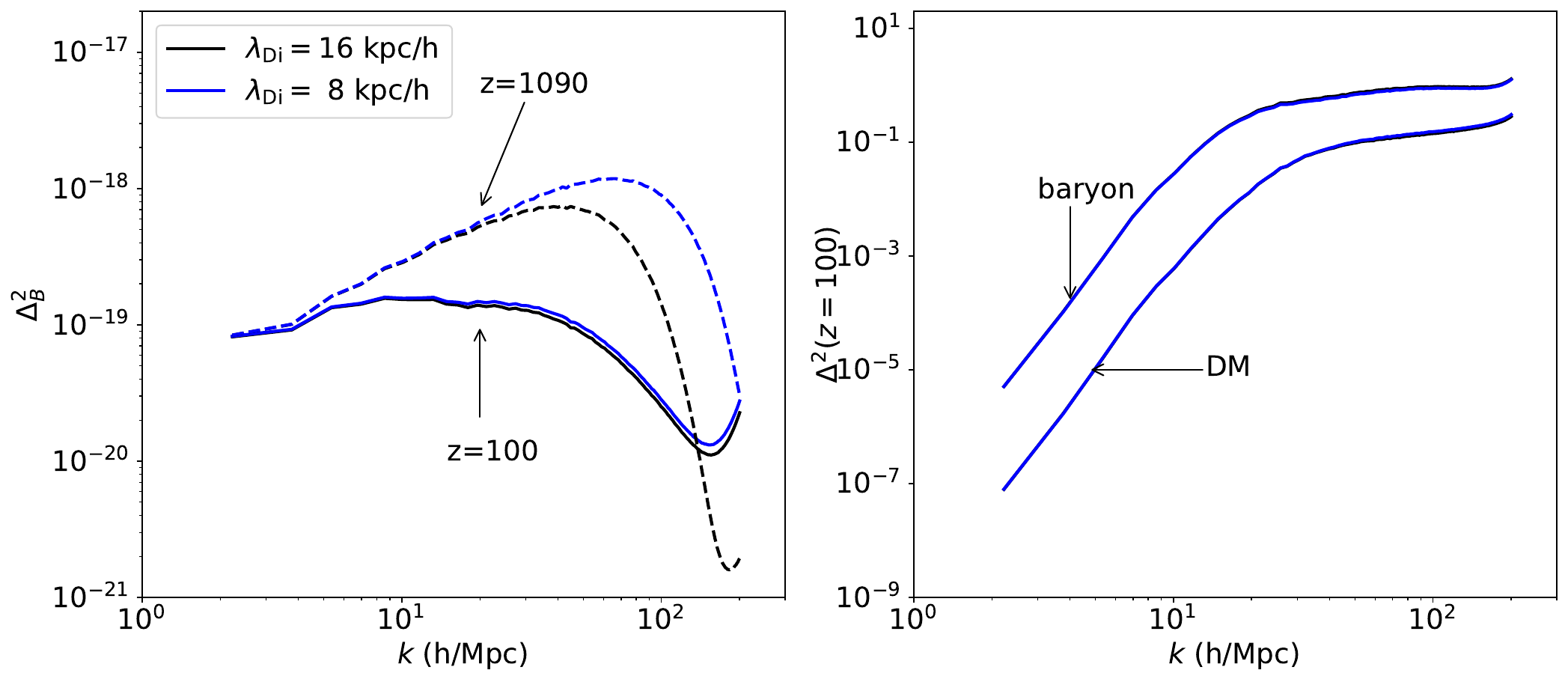}
		\caption{\textbf{Left:} Dimensionless magnetic field power spectrum for $B_{\rm 1Mpc}=0.2$ nG and $n_B=-2$ in the units of G${}^2$. The solid lines show spectra at $z=100$, while the dashed lines show spectra at $z=1090$. The black and blue lines are for simulations with $\lambda_{\rm Di}=16$ kpc$/h$ and $8$ kpc$/h$, respectively. All other simulation parameters are kept the same. \textbf{Right:} Dimensionless baryon and dark matter power spectrum for the two simulations. The colour code is same as in the left panel.}\label{fig:lambda_ic_nB_minus2}
	\end{figure}

	In figure~\ref{fig:lambda_ic_nB_minus2} we compare the power spectra for simulations with $n_{\rm B}=-2$ but with different values of $\lambda_{\rm Di}$. Here $\lambda_{\rm Di}$ is the magnetic damping length scale just before the end of recombination, see discussion around eq.~\eqref{eq:lambda_Di}.
	One can see that changing $\lambda_{\rm Di}$ has negligible impact on the power spectrum at $z=100$.
	
	\section{Numerical issues with scale-invariant fields}\label{sec:scale_invt}
	In this appendix, we show that our modified NGenIC code does not accurately produce magnetic fields for scale-invariant power-spectrum. We believe this might be a general problem for scale-invariant magnetic fields and not just limitations of our specific code.
	
\begin{figure}
		\includegraphics[width=1.00\textwidth]{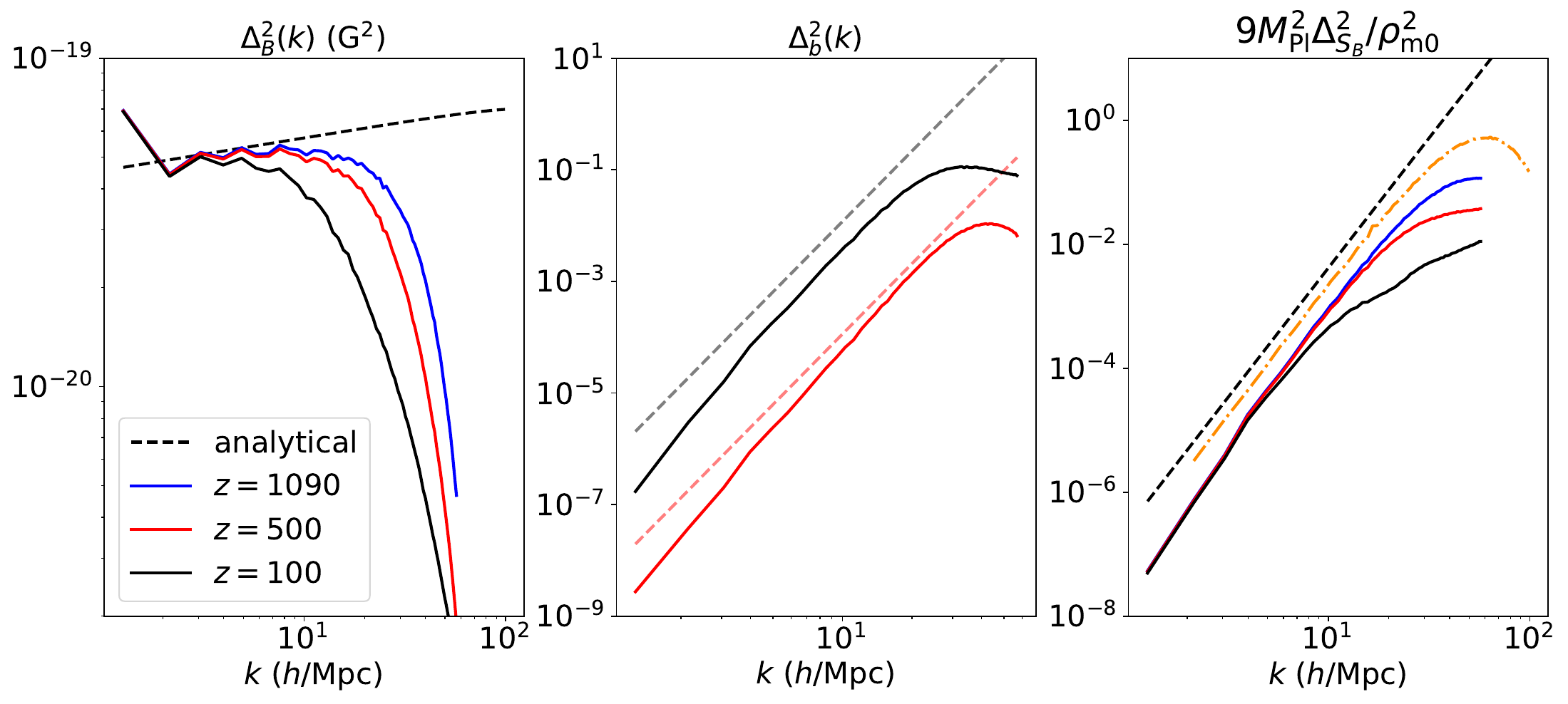}
		\caption{Dimensionless power spectrum of magnetic fields (left), baryons (center), and $3M_{\rm Pl}^2S_B/\rho_{\rm m0}$ (right) at different redshifts for $B_{\rm 1Mpc}=0.678$ nG and $n_{\rm B}=-2.9$. The dashed lines in the panels correspond to the power spectrum computed using linear theory. The orange dot-dashed line in the right panel shows the analytical power spectrum after including IR and UV cut-offs due to the finite box size.}\label{fig:nB_2_9}
	\end{figure}

	In the left and centre panel of figure~\ref{fig:nB_2_9} we show the evolution of PMF and baryon power spectrum for $B_{\rm 1Mpc}=0.678$ nG, $n_{\rm B}=-2.9$, and $\lambda_{\rm D,i}=\infty$. Refer to section~\ref{sec:setup} for more details on how our simulations are set up. The box size of our simulation is 7 Mpc/$h$ and the number of baryon particles is $128^3$.
	
	The dashed lines in the centre panel of figure~\ref{fig:nB_2_9} show the expectation from linear theory, computed using eq.~\eqref{eq:Delta_b}. The numerical power spectrum on large scales is suppressed compared to the linear theory by a factor of around 5 at all redshifts.
	
	In the right panel, the solid lines show the power spectrum of $3M_{\rm Pl}^2S_B/\rho_{\rm m0}$. Note that $S_B\propto \nabla\cdot (\nabla\times B)\times B)$ and thus it is directly determined by the $B$ fields at each snapshot. We show the $S_B$ power spectrum because from linear theory we expect the distribution of density perturbations to be the same as the distribution of $S_B$. One can see that even the numerical power spectrum of $S_B$ deviates from what is expected (black dashed line) by a factor of 5, even though the power spectra of $B$ fields are as expected.
	
	A likely cause for the suppression could be the finite box size used in the numerical method.
	The analytical \( S_B \) power spectrum is calculated by integrating over all \( k \) modes from 0 to infinity (as per eq.~\eqref{eq:Delta_b}).
	For magnetic fields with spectra that are closer to scale invariance, significant contributions come from a wide range of scales.
	However, the finite box size of the simulation limits the contribution to \( S_B \) to only those \( k \) modes that fit within the box.
	
	To test whether this finite box size can account for the observed suppression in the numerical power spectrum,
	we recalculated the analytical power spectrum using eq.~\eqref{eq:Pb_int}, incorporating the exponential suppression of \( P_B \) at large \( k \),
	and restricted the integral to sample only those \( k \) values larger than the smallest wave number in the NGenIC box.
	
	The resulting power spectrum is shown as the dot-dashed orange line in figure~\ref{fig:nB_2_9}.
	From this, it is evident that while the finite box size contributes to the suppression,
	It cannot fully account for the reduced \( S_B \) power spectrum observed in the numerical method.
	
	One can also heuristically see how much the IR cutoff influences the $S_B$ power spectrum. With an IR cutoff, the $S_B$ power spectrum is given by
	\begin{align}\label{eq:Delta_S0}
		\Delta_{S_B}(k)\approx 10^{-4} \left(\frac{k}{\rm Mpc^{-1}}\right)^{2n_{\rm B}+10} \left(\frac{B_{\rm 1 Mpc}}{\rm nG}\right)^4G_{\rm n_{\rm B}}(k_{\rm IR}/k),
	\end{align}
	where $G_{\rm n_{\rm B}}$ is now determined by
	\begin{align}
		G_{\rm n_{\rm B}}(k_{\rm IR}/k)=\int_{k_{\rm IR}/k}^{\infty} dx \int _{-1}^1\frac{dy}{2}x^{n_{\rm B}+2}(1+x^2-2xy)^{n_{\rm B}/2-1}\frac{\left[1+2x^2+4y^4x^2-4y^2x^2-4y^3x+y^2\right]}{\Gamma^2([n_{\rm B}+3]/2)}.
	\end{align}
	In the limit $k_{\rm IR}/k\ll 1$, the above can be approximated as
	\begin{align}
		G_{\rm n_{\rm B}}(k_{\rm IR}/k)=G_{\rm n_{\rm B}}(0)-\frac{4(k_{\rm IR}/k)^{n_{\rm B}+3}}{3(n_{\rm B}+3)\Gamma^2([n_{\rm B}+3]/2)}
	\end{align}
	For $n_{\rm B}=-2.9$, we have $G_{\rm n_{\rm B}}(0)=0.068$ and the second term on the right is $0.022\left(\frac{k_{\rm IR}/k}{0.01}\right)^{0.1}$. Thus, the IR cutoff can at best only affect the $S_B$ power spectrum on the order of 30\%, while we observe almost an order of magnitude suppression in the numerical power spectrum.

	This issue of the suppressed value of $S_B$ power spectrum for scale-invariant fields was also found in Ref.~\cite{Ralegankar:2024ekl}. In that study, it was shown that the suppression factor is robust to changes in grid and box sizes and is only a function of $n_{\rm B}$. The suppression is reduced as $n_{\rm B}$ increases, ranging from a factor of 5 for $n_{\rm B}=-2.9$ to a factor of 1.1 for $n_{\rm B}=-2.0$.

	\section{Resolution check for $n_{\rm B}=2$}\label{sec:convergence}
	\begin{figure}
		\includegraphics[width=1.00\textwidth]{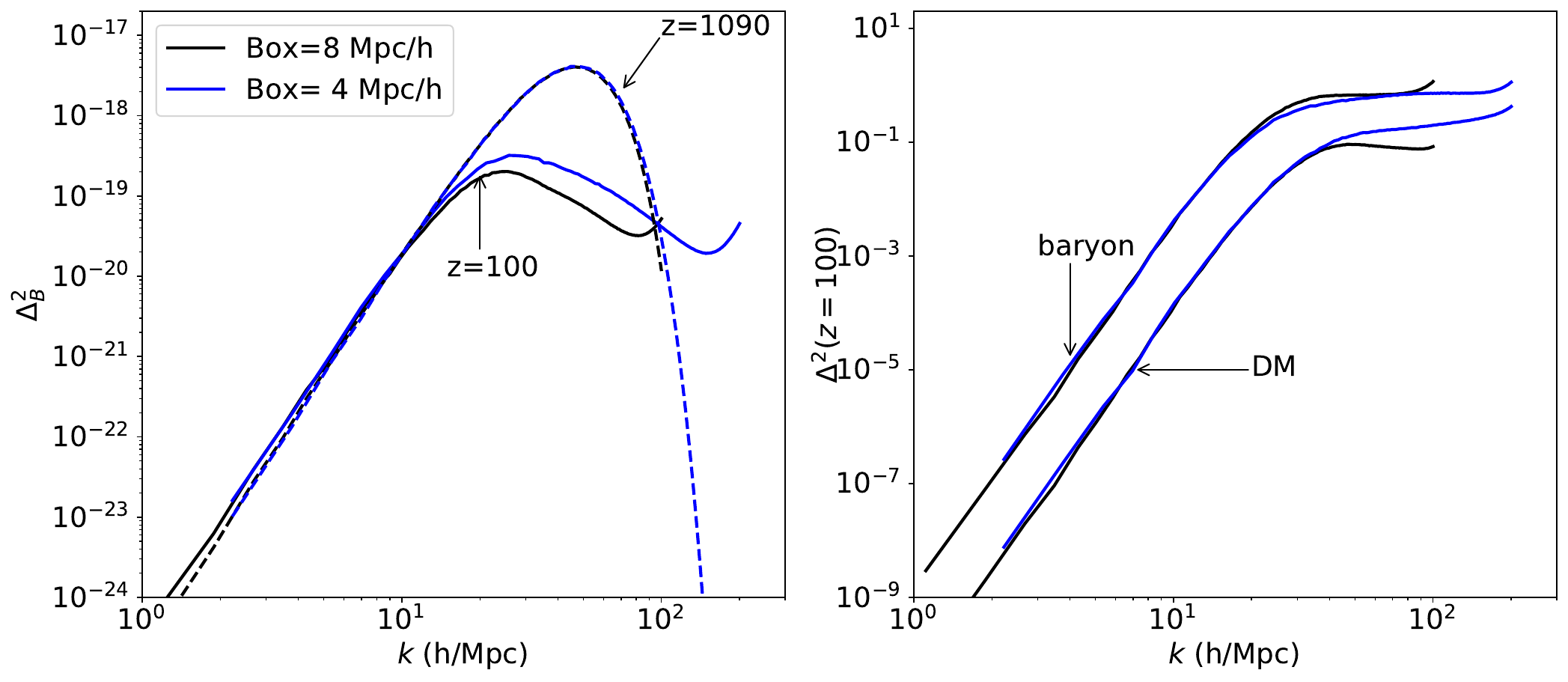}
		\caption{\textbf{Left:} Dimensionless magnetic field power spectrum in the units of G${}^2$. The solid lines show spectra at $z=100$, while the dashed lines show spectra at $z=1090$. The black and blue lines are for simulations E and F (see Table~\ref{table:simparams}), respectively. Both simulations have same configurations except the box sizes. \textbf{Right:} Dimensionless baryon and dark matter power spectrum for different simulations. The colour code is same as in the left panel.}\label{fig:resolution_nB_plus2}
	\end{figure}
	
	In figure~\ref{fig:resolution_nB_plus2} we compare the power spectra for simulations with $n_{\rm B}=2$ but with different resolutions. We see that lower resolution leads to more numerical damping of magnetic fields on small scales, which was also seen for $n_{\rm B}=-2$ case in section~\ref{sec:nB_minus2_res}. The baryon power spectrum on small scales saturates to a constant and that value is independent of resolution. However, dark matter power on small scales is found to have mild sensitivity to the resolution of the simulation.

	\acknowledgments
	The authors thank Mak Pavi\v{c}evi\'{c}, Romain Teyssier, Adrianne Slyz, Julien Devriendt, Mahsa Sanati, Sergio Martin-Alvarez, and Takeshi Kobayashi for useful conversations. We are thankful to the community for developing and maintaining software packages extensively used in our work, namely:  \texttt{matplotlib} \citep{matplotlib}, \texttt{numpy} \citep{numpy} and \texttt{scipy} \citep{scipy}. MV acknowledges support by the Italian Research Center on High
	Performance Computing, Big Data and Quantum Computing (ICSC), project
	funded by European Union - NextGenerationEU - and National Recovery and Resilience Plan (NRRP) - Mission 4 Component 2, within the activities of Spoke   3, Astrophysics and Cosmos Observations, and by the INFN Indark Grant. All the simulations presented in this work have been run on the Ulysses supercomputer at SISSA.	
	
	\bibliographystyle{utphys}
	\bibliography{references}
	
\end{document}